\documentclass[10pt,journal,compsoc]{IEEEtran}
\pdfoutput=1

\usepackage{cite}

\usepackage{caption}
\usepackage{graphicx}
\usepackage{float} 
\usepackage{subcaption}

\usepackage{amssymb}
\usepackage{amsmath}
\usepackage{epstopdf}
\usepackage{color}
\usepackage[ruled,linesnumbered]{algorithm2e}
\usepackage{bm}

\usepackage{url}
\usepackage{booktabs}
\usepackage{etoolbox}

\usepackage{array}

\makeatletter
\patchcmd{\@makecaption}
  {\scshape}
  {}
  {}
  {}
\makeatother

\begin{document}

\title{Dependent Task Offloading in Edge Computing \\ Using GNN and Deep Reinforcement Learning}

\author{Zequn~Cao, Xiaoheng~Deng

        
}

\markboth{}
{Shell \MakeLowercase{\textit{et al.}}: Bare Advanced Demo of IEEEtran.cls for IEEE Computer Society Journals}

\IEEEtitleabstractindextext{
\begin{abstract}
  Task offloading is a widely used technology in Mobile Edge Computing (MEC), which declines the completion time of user task with the help of resourceful edge servers. 
  Existing works mainly focus on the case that the computation density of a user task is homogenous so that it can be offloaded in full or by percentage. However, various user tasks in real life consist of several inner dependent subtasks, each of which is a minimum execution unit logically. Motivated by this gap, we aim to solve the Dependent Task Offloading (DTO) problem under multi-user multi-edge scenario in this paper.
  We firstly use Directed Acyclic Graph (DAG) to represent dependent task where nodes indicate subtasks and directed edges indicate dependencies among subtasks. Then we propose a scheme based on Graph Attention Network (GAT) and Deep Reinforcement Learning (DRL) to minimize the makespan of user tasks. To utilize GAT efficiently, we put the training of it on resourceful cloud in unsupervised style due to the numerous data and computation resource requirements. In addition, we design a multi-discrete Action space for DRL algorithm to enhance the applicability of our proposed scheme.
  Experiments are conducted on broadly distributed synthetic data. 
  The results demonstrate that our proposed approach can be adapted to both simple and complex MEC environments and outperforms other methods.

\end{abstract}

}

\maketitle

\IEEEdisplaynontitleabstractindextext
\IEEEpeerreviewmaketitle

\section{Introduction}
\IEEEPARstart{W}{ith} the development of the mobile Internet, smart
devices have become an indispensable part of people's life.
The progress of underlying technologies, such as sensor technology 
and wireless network technology, enables all kinds of traditional devices 
to be connected to the Internet, making us enter the era of 
the Internet of Everything (IoE).
Since Google put forward the concept of cloud computing in 2008,
cloud computing was gradually accepted and introduced into 
the mobile environment, which breaks through the resource limitations of 
smart devices and provides highly demanding applications for users\cite{cloud2008}.
Cloud computing provides abundant applications and services in on-demand way, 
regardless of the user's location and the type of smart devices\cite{survey2021}.
However, the centralized processing model of cloud computing has been hard to satisfy the higher requirement
of latency, energy consumption, application performance,
and reliability due to the development of IoE. 
Fortunately, a new computing paradigm — Mobile Edge Computing (MEC) is proposed as a 
powerful supplement to cloud computing.
MEC can better meet the higher requirements by deploying edge server (ES) closer to user equipment (UE) geographically
and integrating distributed resources in edge.

Unlike cloud computing, which theoretically have unlimited resources, 
edge computing must improve Quality of Experience (QoE) of users under resource constraints.
Therefore, how to efficiently utilize edge resources is essential in MEC.
Task offloading is one of the most important technologies of
resource scheduling in MEC\cite{survey2021}. 
In recent years, task offloading in MEC has attracted the attention of academic and industrial
circles. 
By deciding whether a computation-extensive
user application (referred as task) offloaded to the edge to process, this technology can significantly reduce processing delay and
save energy consumption at UE. Selecting an appropriate task to offload to a specific ES 
could be hard, since many factors need to be considered such as task profiles, UE profiles, ES profiles and 
communication status.
In terms of granularity, task offloading problem can be divided into two types:
binary offloading and partial offloading. Binary offloading also known as 0-1 offloading,
means that the whole computation task processed either locally or offloaded to edge servers.
Partial offloading gives a flexible task data partitioning compared to binary offloading. 
By adjusting offloading ratio, it utilizes both local and server resources, further improving QoE. 

However, this partial offloading is hard to realize in practice, because it may
break the code composition structure of task.
Dividing a complex task into parts requires taking into
account the internal dependencies of the task. In many cases, a task 
can be divided into a number of subtasks which depend on each other to complete the entire task. 
Using a series of abstract data to present task makes dependency information lost, 
which may have a great impact on offloading decision-making, resulting in poor QoE.

We use Directed Acyclic Graph (DAG) to represent complex task \cite{dagTC2021,dagIOT2022}. Nodes in DAG
represent application subtasks, and directed edges represent dependencies among subtasks. Each node is a minimum executable and offloading unit. 
In this paper, we aim to schedule all these nodes to local UE or remote ES without breaking dependencies among them in a minimum task completion time.
In this case,
this Dependent Task Offloading (DTO) problem in MEC can be regarded as a type of DAG task scheduling in
heterogeneous computing system\cite{HEFT}.
Efficiently scheduling DAG leads to a hard algorithmic problem whose 
optimal solution is intractable\cite{OSDI2016,sigcom}. Many existing studies 
developed heuristic or approximation algorithms. However, they were either limited by 
the familiar specific expert knowledge or difficult to scale up. 

Deep Reinforcement Learning (DRL), which combines Reinforcement Learning (RL) with
Deep Neural Network (DNN), has demonstrated its powerful ability in many decision-making fields, such as
Atari video games\cite{DQN}, robotics\cite{DDPG}, autonomous driving\cite{ITSC2019}, etc. 
Lots of existing works used DRL for task offloading\cite{inTII2022, inTOC2020, inTVT2019qiu, inTC2020},
however, they all aimed to solve traditional binary offloading or partial offloading, which ignores the intrinsic task dependency.
A few DRL-based studies considered the dependencies among subtasks \cite{dagTC2021,dagIOT2022}, nevertheless, they didn't make full advantage of DAG and only proposed DTO scheme under small-scale MEC scenario. 

We propose a DRL-based DTO scheme (DTODRL) combing Graph Attention Network (GAT)\cite{GAT} and Proximal Policy Optimization (PPO)\cite{PPO} for multi-user multi-edge scenario.
To capture the graph information and features of nodes and edges, GAT is used to encode DAG into State vector that DRL can receive. PPO provides better sample efficiency, stable training and avoids stucking in the local optimal compared to other DRL algorithm. 
The DTODRL can efficiently learn a DTO scheduler and minimize the completion time of tasks.

The major contributions of this paper are summarized as follows:
\begin{enumerate}
  \item We develop an DRL-based DTO scheme (DTODRL),
  which considers the data and logical dependencies among subtasks in user application. We design a State space with comprehensive representation of system information, a multi-dimension Action space that fits the process of scheduling and goal-oriented Reward function.
  The DTODRL is capable of multi-user multi-edge scenario and
  adaptive to a larger scale. 
  \item To make full advantage of DAG, we use GAT to encode it as part of State representation for DRL. GAT encoder is pretrained in unsupervised style on resourceful cloud and decoupled from edge layer scheduler training, which makes scheduler training steadily and faster.
  \item Experiments are conducted on widely distributed synthetic data. The results show that the DTODRL outperforms other methods under both simple and complex MEC scenario.
\end{enumerate}

The rest of this paper is organized as follows.
The related work is introduced in Section 2. Section 3 describes designed MEC system model. Section 4 gives the detail of our proposed scheme. Experiment results are 
presented and analyzed in Section 5. Section 6 conclude this paper and discuss future works.

\section{Related Work}
\subsection{Task Offloading}
In recent years, task offloading has become a hotspot research area. \cite{survey2021} classified resource scheduling
in MEC, in which task offloading is one of the most important components. \cite{survey2017} gave a thorough
survey on architecture and task offloading. 
\cite{inTPDS2019hong} adopted a game-theoretic approach to  
achieving Quality of Service (QoS)-aware task offloading for the
industry IoT-edge-cloud computing model.
In \cite{inIOT2021yang}, offloading node selection strategy was formulated as an MDP, and solved by
the value iteration algorithm (VIA) under multi-user multi-edge scenario.
\cite{inTVT2019} proposed a collaborative task offloading and resource allocation optimization (CCORAO) for cloud assisted MEC in vehicular networks.
\cite{inIOT2021wu} proposed an online offloading algorithm based on the Lyapunov optimization for blockchain-enabled IoT-edge-cloud offloading architecture.

There are also many DRL-based task offloading studies\cite{inTII2022,inTOC2020,inTVT2019qiu,inTC2020}.
\cite{inTII2022} used deep Q-network (DQN)\cite{DQN} to solve the problem of digital twinning-empowered Internet of Vehicles (IoV) in MEC.
\cite{inTOC2020} proposed a scheme that combines Q-network and actor-critic algorithm to improve the computational performance in MEC.
\cite{inTVT2019qiu} introduced an adaptive genetic algorithm into the exploration of DRL to avoid useless exploration and speed up the convergence.
\cite{inTC2020} studied the offloading problem without information sharing and formulate it as a multiagent partially observable MDP, then solve it by a policy gradient DRL.

Above studies all belong to binary offloading or ordinary partial offloading and have their limitations.
DTO problem has gradually become a new hotspot in task offloading\cite{dagTC2021,dagIOT2022, dagTPDS2021}. 
\cite{dagTPDS2021} studied the problem of offloading dependent tasks with service-caching (ODT-SC) and designed a convex programming based algorithm (CP) to solve this problem. But the proposed scheme requires expert knowledge.
\cite{dagTC2021} converted raw DAG into a sequence, then use an S2S neural network\cite{s2s} to represent the policy and the value function of DRL under single-user single-edge scenario.
However, inserting S2S neural network into DRL framework makes DRL training slow, unstable and hard to scale up for a complex scenario.
\cite{dagIOT2022} introduced Graph Convolutional Network (GCN)\cite{GCN} to get nodes' embedding vectors as State before input into DRL framework. They generated node embedding vector by transferring the information from back nodes to front nodes, which didn't fit the scheduling forward process.
Besides, coupling GCN and DRL output layer together during training makes it hard to meet the requirement of real-time scheduling.

\subsection{Representation Learning for DRL}
Using DRL to solve problems in MEC is widely studied, however, as in other areas, it is also plagued by common problems in DRL like inefficient sampling, dimensional disasters, and poor interpretability. Existing works\cite{stooke2021decoupling, lan2022generalization,tpami23,iral18,pmlr19,pmlr22} have shown that representation learning alleviates these problems to some extent and enhances the robustness and applicability of DRL.
\cite{iral18} attributed the large interaction requirements to learn useful, general control policies for robotics to the fact that a state representation needs to be learned as a part of learning control policies, which can only be done through fitting expected returns based on observed rewards. While the reward function provided information on the desirability of the state of the world, it didn't necessarily provide information of how to distill a good, general representation of that state from the sensory observations.
\cite{stooke2021decoupling} introduced an unsupervised learning called Augmented Temporal Contrast (ATC), which trains a convolutional encoder to associate pairs of observations separated by a short time difference, under image augmentations and using a contrastive loss\cite{moco}. They showed that training the encoder exclusively using ATC matches or outperforms end-to-end RL in most environments.
\cite{tpami23} proposed an algorithm called masked contrastive representation learning for pixel-based RL (M-CURL) where the states were raw video frames, which took the correlation among consecutive inputs of images into consideration. They trained CNN encoder and auxiliary Transformer encoder via contrastive learning where the reconstructed features should be similar to the groud-truth ones while dissimilar to others.

Thanks to the development of graph representation learning\cite{dgi,udgsl,vgae,gate} and the ability of aggregating widely distributed data of Cloud-Edge-User MEC framework, we pretrain a GAT encoder in unsupervised style in resourceful cloud layer, which aims to capture comprehensive graph-based state representation.
Inspired by graph scheduling in other fields\cite{HEFT,OSDI2016,sigcom} and DRL in task offloading\cite{inTII2022,inTOC2020,inTVT2019qiu,inTC2020,dagTC2021,dagIOT2022}, we then design a multi-dimension Action Space DRL to simulate general DTO process. Our proposed DTODRL can be adaptive to various scenarios and get near-optimal solutions.

\begin{figure}[t] 
  \centering
  \includegraphics[width=0.4\textwidth]{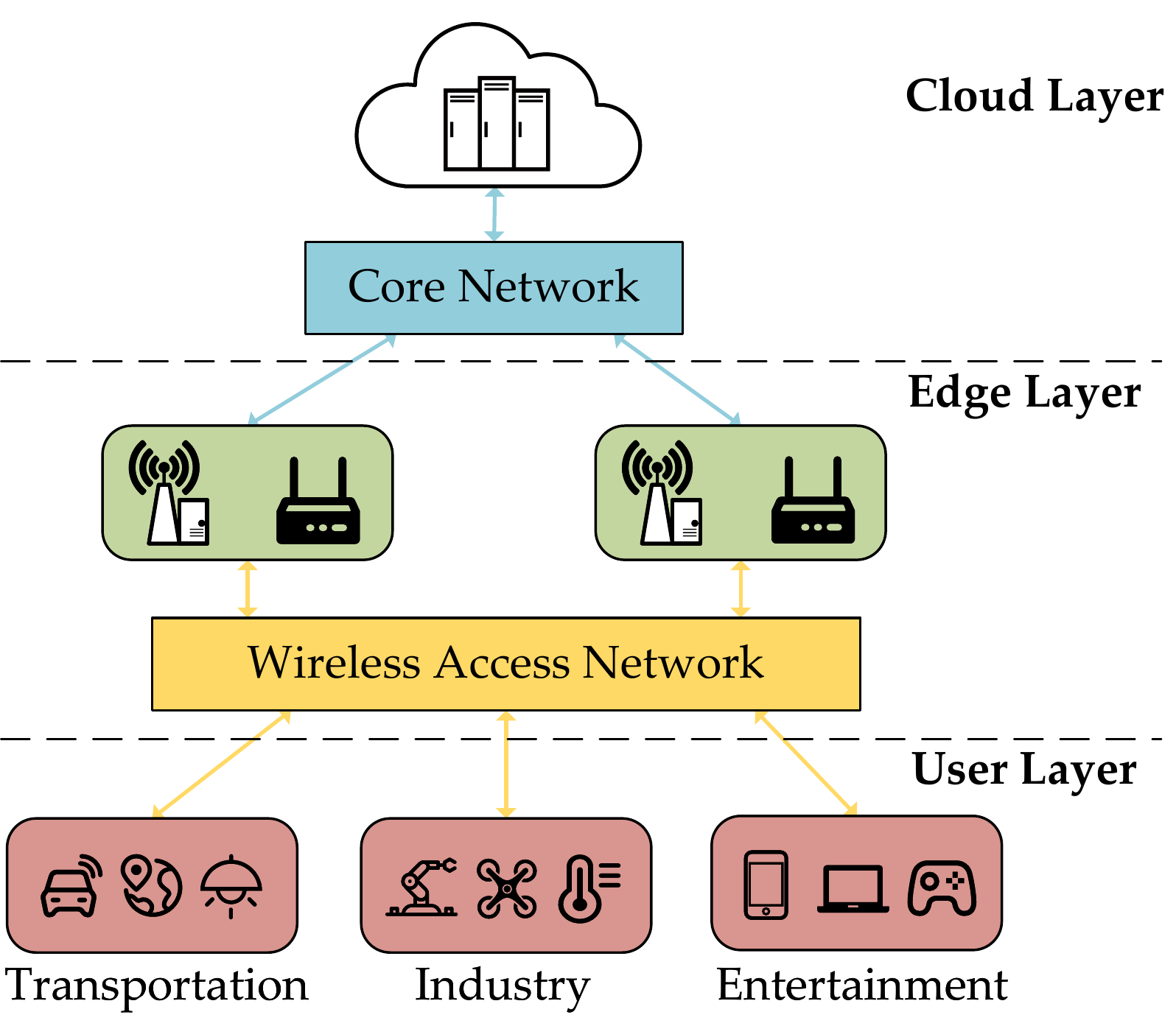}
  \caption{General Cloud-Edge-User MEC framework.}
  \label{fig:1}
  \vspace{-0.35cm}
\end{figure}

\section{System Model}
This section firstly introduces the framework of DTO problem and the detail of DTO process in our designed system, 
then presents related calculation formulas. In the end, the optimization goal will be given.

\subsection{Overview Design}
Fig. \ref{fig:1} shows a typical Cloud-Edge-Server MEC framework. This framework consists of three layers,
namely cloud layer, edge layer and user layer. Mobile UE communicate with ES by wireless access
network, while edge servers are connected with cloud datacenter by core network (CN). 
Cloud servers have massive computation resources, but limited by bandwidth congestion and heavy load on CN. 
UEs have the weakest computing capabilities, but they are the generation source of computation tasks. 
ESs are deployed closer to UEs, making the communication overhead between them is lower than that between 
UEs and cloud directly. 

In this work, we consider offloading location is only between ES and
UE, i.e., UE not only is the producer of task but also can be used to execute nodes. Cloud layer is used to pretrain a GAT encoder that requires enormous computational resources, which will be discussed in section 4.
In this case, the set of ESs and UEs are denoted by $ ES=\left\{ es_1,es_2,...,es_m,...,es_M \right\}  $ 
and $ UE=\left\{ ue_1,ue_2,...,ue_k,...,ue_K \right\} $.
Task of $ue_k$ can be represented as a DAG $G_k=\left( \nu_k,\varepsilon_k \right) $, where $\nu_k$ is the set
of nodes also indicates subtasks of $ue_k$, and $\varepsilon_k$ is the set of directed edges between nodes also indicates dependency between subtasks of $ue_k$.
All nodes are indexed in topological order.
We add \emph{Start} node and \emph{End} node to the front and the last of all DAGs respectively.
\emph{Start} node and \emph{End} node have no practical executable data, they are only used to sign the initial and completed status. We use $N_k$ to denote the number of nodes of $G_k$, thus, there are $N_k-2$ nodes are executable.
Nodes in DAG task can be executed on its own UE or any ES, and for each node $\nu_k^i$ in $ue_k$, its \emph{available offloading location (AOL)} set is marked as $AOL_{\nu_k^i}=AOL_k=\left\{ ue_k,es_1\,\,,...,es_M \right\} $.
We use $a_k^i$ to indicate the offloading location of node $\nu_k^i$, and $a_k^i\in AOL_k$. 

\begin{figure}[h] 
  \centering
  \includegraphics[width=0.48\textwidth]{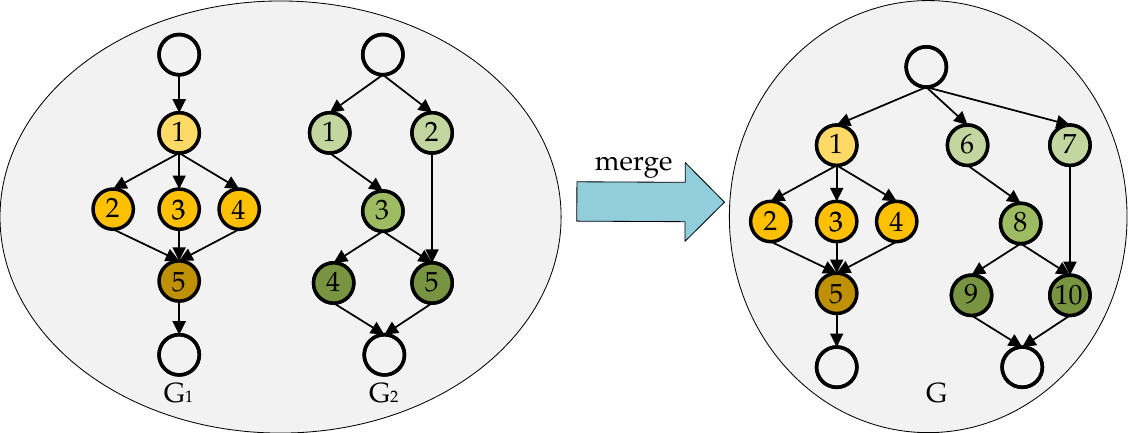}
  \caption{
    We use different color of node to indicate different UE's task. As the DAG goes deeper, the color of nodes are also getting darker. Each DAG has a private \emph{Start} node and \emph{End} node, while the merged DAG has a public \emph{Start} node and several private \emph{End} nodes for each UE's task.}
  \label{fig:merge}
\end{figure}

Each task of single-user is represented as a DAG with extra \emph{Start} node and \emph{End} node, then, we connect all DAGs by merging their \emph{Start} node and reindex nodes for the multi-user scenario.
Fig. \ref{fig:merge} shows how to merge DAGs of multi-user tasks. The public \emph{Start} node still signs the initial status of scheduling, while each \emph{End} node indicates the completed status of its corresponding task. By doing so, the parallelism between tasks remains unchanged, the scheduling of any node will only affect the scheduling of its own task, and the merged graph can be directly used as input to GNN. For convenience, the rest of the paper will be discussed based on the merged DAG, and some notations will also change accordingly (from $\nu_k^i$, $a_k^i$ for $G_k$ to $\nu_i$, $a_i$ for $G$). In particular, the original DAG of single-user scenario and the merged DAG of it are the same. The main notations in this paper are listed in Table \ref{table:1}. 

\begin{table}[htbp]
  \renewcommand\arraystretch{1.35}
  \caption{Main Notation Definition in This Paper.}
  \vspace{-0.2cm}
  \begin{center}
    \begin{tabular}{c|p{6.5cm}}
      \hline

      \hline
      \textbf{Notations} & \textbf{Description} \\    
      \hline
      $UE$ & The set of user equipment.\\
      \hline
      $ES$ & The set of edge servers.\\
      \hline
      $K$ & The number of user equipment.\\
      \hline
      $M$ & The number of edge servers.\\
      \hline
      $U$ & Total number of user equipment and edge servers.\\
      \hline
      $G_k$ & The DAG task of ${ue}_k$. \\
      \hline
      $G$ & The merged DAG. \\
      \hline
      $\nu_k^i$ & The $i$th node of $G_k$. \\
      \hline
      $\nu_i$ & The $i$th node of $G$. \\
      \hline
      $N_k$ & The number of nodes of $G_k$.\\
      \hline
      $N$ & The sum number of nodes of all DAGs.\\
      
      \hline
      $f_k^{ue}$ & The CPU clock speed of ${ue}_k$.\\
      \hline
      $f_m^{es}$ & The CPU clock speed of ${es}_m$.\\
      \hline
      ${tr}^s$ & The transmission rate between edge servers. \\
      \hline
      ${tr}^l$ & The transmission rate between edge server and user equipment. \\
      \hline
      ${tr}_{i,j}$ & The transmission rate between locations of $\nu^i$ and $\nu^j$,
      either equal to $r_s$ or equal to $r_l$. \\
      
      \hline
      $C_i$ & The required CPU cycles for executing $\nu_i$.\\
      \hline
      $D_i$ & The required data of $\nu_i$ to upload to execution location.\\
      \hline
      $Pred(\nu_i)$ & The set of $\nu_i$'s predecessor nodes.\\
      \hline
      $Succ(\nu_i)$ & The set of $\nu_i$'s successor nodes.\\
      \hline
      $D_{j,i}$ & The transmission data from $\nu_j$ to $\nu_i$.\\
      
      \hline
      $a_i$ & The offloading location of $\nu_i$.\\
      \hline
      $Proc(a_i)$ & The set of $a_i$'s processors.\\

      \hline
      $T^{ul}_{i}$ & The upload time of $\nu_i$.\\
      \hline
      $T^{tr}_{j,i}$ & The transmission time from $\nu_j \in Pred(\nu_i)$ to $\nu_i$.\\
      \hline
      $T^{ex}_{i}$ & The execution time of $\nu_i$.\\
      \hline
      $T^{dl}_{i}$ & The download time of $\nu_i$.\\

      \hline

    \end{tabular}
  \end{center}
  \label{table:1}
\end{table}

\subsection{Node Offloading}
The completion of each node is composed of four parts: \emph{Upload}, \emph{Transmission}, \emph{Execution} and \emph{Download}, 
while the completion of a task is not the sum of all its nodes' completion time due to the dependency and parallelism among nodes.
Next, offloading model for single node and graph scheduling model for whole task are described respectively.

\begin{figure*}[t] 
  \centering
  \includegraphics[width=0.95\textwidth]{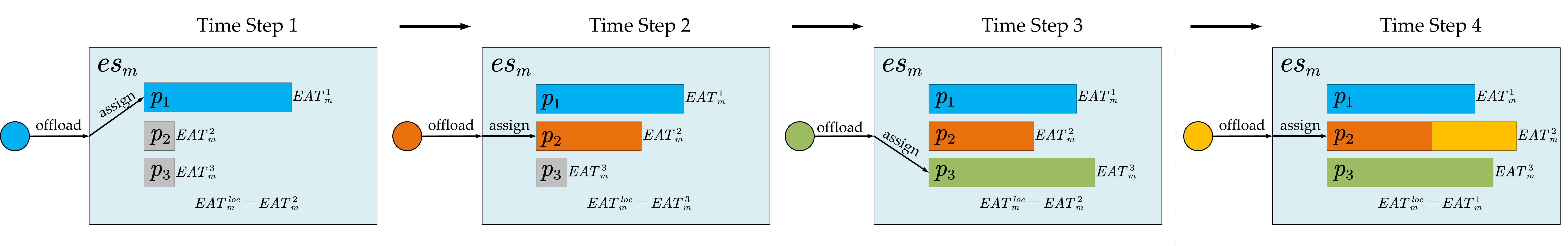}
  \caption{How \emph{EAT} of $es_m$ changes when 4 nodes offloaded to it successively. In this case, $es_m$ has 3 processors: $p_1$, $p_2$ and $p_3$. In the initial state, all processors' \emph{EAT} of $es_m$ is set to 0. The color of processor's rectangle indicates its receiving nodes.
  The length of processor's rectangle represents how long it takes the processor to execute the node, i.e. $T^{ex}$.}
  \label{fig:assign}
\end{figure*}

\subsubsection{Upload Model}
In this work, we consider a block fading channel, where the fading coefficient remain unchanged(i.e., the transmission rate is fixed)\cite{dagTC2021}.
The transmission rates between any UE and any ES are the same denoted as ${tr}^l$, and the transmission rates between ESs are the another same value denoted as ${tr}^s$.
Upload time is only calculated for those nodes who are offloaded to ES. If node $\nu_i$  is decided to be executed
on its own UE, the upload time $T^{ul}_{i}=0$.
Otherwise, if it is offloaded to $es_m$, then we have: 
\begin{equation}
  T^{ul}_{i}=\frac{D_i}{{tr}^l},
\end{equation}
where $D_i$ indicates how much necessary program data needs to be uploaded for the execution of $\nu_i$. 

\subsubsection{Transmission Model}
Transmission time is the time of transmitting necessary data from $\nu_j \in Pred(\nu_i)$ to activate $\nu_i$'s execution, 
where $Pred(\nu_i)$ is the set of predecessor nodes of $\nu_i$. No node can start execution before all of its predecessor nodes have been executed due to the dependency between them. But $\nu_i$ can be uploaded to any ES
before it receives data from its predecessor nodes, that is, the execution of $\nu_j \in Pred(\nu_i)$ and the upload of $\nu_i$ can be parallel, considering the structure of task is known in advance and the execution code data 
of nodes is logically decoupled from each other. 
If $\nu_i$ is scheduled to the same location as $\nu_j\in Pred(\nu_i)$,
the transmission time inner a same location is negligible.
Therefore, the transmission time from $\nu_j$ to $\nu_i$ can be divided into two cases:
\begin{equation}
  T^{tr}_{j,i}=\begin{cases}
  \frac{D_{j,i}}{{tr}_{j,i}}, a_i\neq a_j\\
  0, a_i = a_j\\
  \end{cases},
\end{equation}
where ${tr}_{j,i}$ is the transmission rate between locations of $\nu_j$ and $\nu_i$ and $D_{j,i}$ indicates the transmission data size from $\nu_j$ to $\nu_i$. 


\subsubsection{Execution Model}
We consider ESs have several parallel processors with the same computation ability denoted as $f^{es}$. UEs have only one processor and the CPU clock speed of all UEs are the same denoted as $f^{ue}$.
Each ES can continuously receive nodes from any UE, but it can only execute as many nodes as its parallel processors at the same time. The execution for nodes in a same location follows the first in first out (FIFO) manner.
Each node is either executed locally or on an ES. Thus, the execution time of $\nu_i$ can be calculated in two cases:
\begin{equation}
  T^{ex}_{i}=\begin{cases}
    \frac{C_{i}}{f^{ue}}, a_{i}\,=0\\
    \frac{C_{i}}{f_{m}^{es}}, a_{i}\,\neq 0\\
  \end{cases},
  \end{equation}
where $C_{i}$ is the required CPU cycles for executing $\nu_i$.

\subsubsection{Download Model}
Not all intermediate results of executable nodes (except
\emph{Start} node and \emph{End} node) need to be returned to UE.
\emph{Download} only happens for those 
executable nodes who connect to \emph{End} node directly, i.e., $\nu_i\in Pred(End)$. Besides, only those nodes that are 
uploaded to ES need to return their results back to UE. The download time $\nu _{i}$ is obtained by
\begin{equation} 
  T^{dl}_{i}=\begin{cases}
    \frac{D_{i,End}}{{tr}_l}, Flag\left( \nu _{i} \right)=1\\
    0, Flag\left( \nu _{i} \right)=0
  \end{cases},
\end{equation}
where $Flag\left( \nu _{i} \right)$ is a binary variable used to represent whether $\nu _{i}$ is offloaded and whether $\nu_i\in Pred(End)$, $D_{i,End}$ indicates the return data size from $\nu_i$. Actually, download can be regarded as a special case of transmission.

\subsection{Scheduling Model}
As we mentioned above, it is necessary to consider not only each single node, but also their parallelism in the DAG. No matter how fast nodes in DAG can be completed on average, the completion of whole task still be limited by the last done node, i.e., \emph{End} node. The completion of single node is composed of four parts, but its completion time is not the order sum of all four parts due to the competition for network and computing resources. We consider uplink waiting queue, downlink waiting queue, transmission waiting queue and execution waiting queue to limit the competitor for resources. In this paper, we assume each UE holds an transmission waiting queue and each ES holds $K+M-1$ transmission waiting queue. Uplink waiting queue and downlink waiting queue belong to UE and ES, respectively.
Each processor corresponds to an execution waiting queue.
Once a node is scheduled to offload to ES, it will be sent into the uplink waiting queue, and then wait for the previous nodes to finish uploading. 

In this work, each node has \emph{Start Time (ST)} and \emph{Finish Time (FT)} attributes which are the wall clock time of begin and end of node's completion, respectively. We use \emph{ Estimated Available Time (EAT)} to indicate the available time of several resources (waiting queues).
For \emph{Start} node, \emph{ST} and \emph{FT} are all initialized to 0. For $\nu_i$, its \emph{ST} is determined by three items: 1) the \emph{EAT} of execution location $a_i$; 2) the \emph{EAT} of upload channel from local to execution location $a_i$; 3) max of the transmission time from its predecessor nodes' offloading location to $a_i$, which is obtained by:
\begin{small}
  \begin{equation}
    ST_{i} =\max \left\{EFT^{loc}_{a_i}, EFT^{ul}_{i,{a_i}},\mathop{\max}\limits_{\nu _{i}\in Pred(\nu _{i})}\left\{FT_{j}+T_{j,i}^{tr}\right\}\right\},
  \end{equation}
\end{small}
where the execution in $a_i$, the upload of $\nu_i$ and the transmission from $Pred(\nu _{i})$ are parallelized.
\emph{EAT} of UE or ES is determined by \emph{EAT} of their processors, thus $EAT^{loc}_{a_i}=\mathop{\min}\limits_{p\in Proc(a_i)}\left\{EAT_{a_i}^p\right\}$, where $Proc(a_i)$ is the set of $a_i$'s processors. 
Once node $\nu_i$ is scheduled to a specific location $a_i$, it will be assigned to the FIFO waiting queue of the processor with minimal \emph{EAT}, then, \emph{EAT} of this processor is updated by: $EAT_{a_i}^p = EFT_{a_i}^p + T_{i}^{ex}$. Fig. \ref{fig:assign} shows how \emph{EAT} of $es_m$ changes when it receives 4 nodes successively. 
Then, $\nu_i$'s \emph{FT} is calculated by:
\begin{equation}
  FT_{i}= ST_{i}+T_{i}^{ex}.
\end{equation}
Specially, $FT_{End_k}=ST_{End_k}$ is the \emph{Actual Finish Time (AFT)} of $ue_k$'s task, which illustrates the completion time of it.

\subsection{Problem Formulation}
An object of task offloading is to improve QoE.
In this case, we measure multi-user QoE by using mean \emph{AFT} of users, which directly
reflects how long time the whole completion takes in a given offloading plan.
Therefore, the optimization goal in this paper as follows:
\begin{equation}
  \min \frac{\sum_{k=1}^K{AFT_k}}{K}.
\end{equation}
Most of existing task offloading works \cite{inTC2021,dagTC2021,dagIOT2022} took energy consumption into account, while some \cite{inTMC2022} of them  didn't.
This work doesn't consider energy consumption for the following reasons.
In our model, minimizing mean \emph{AFT} results in low energy consumption, since our proposed method reduces the pressure on local execution, which is the main contributor of high energy consumption. Besides, considering it or not doesn't affect the design of our proposed scheme that can be extended according to practical requirements for optimization goal.

\section{The DTODRL Scheme}
This section presents the proposed DTODRL scheme in detail.
Firstly, we will introduce the design of GAT auto-encoder. Then, the detail of the DTODRL are presented, and how it works for DTO problem are described. Finally, the algorithm are illustrated step by step.

\begin{algorithm}	
  \caption{The Training Process of Unsupervised GAT Auto-Encoder}
  \label{algo:encoder}
  \textbf{Initialize} DAG datasets $B$ \\		
  \textbf{Initialize} k-layers GAT encoder $Q$ and k-layers GAT decoder $\hat{Q}$  \\		
  \For{each epoch}{
    \For{each DAG $G$}{
      Read $G$'s node features matrix $X$, $G$'s adjacency matrix $adj$ from $B$ \\
      $\bm{h^0}=X$ \\
      \For{$l=1, ..., L$}{ 
        $\bm{h^{(l)}} \leftarrow layer_{l}(\bm{h^{(l-1)}}, adj)$
      }
      $\bm{\hat{h}^{(0)}} = \bm{h^{(L)}}$ \\

      \For{$l=1, ..., L$}{
        $\bm{\hat{h}^{(l)}} \leftarrow \hat{layer}_{l}(\bm{\hat{h}^{(l-1)}}, adj)$
      }
      $\bm{h^{\prime}} = \bm{\hat{h}^{(L)}}$ \\
      Compute Feature Loss according to Eq. \ref{eq:featureloss} \\ 
      Compute Structure Loss according to Eq. \ref{eq:structureloss} \\ 
      Update $Q$ and $\hat{Q}$ with (Feature Loss + Structure Loss) \\ 
    }
    
  }
  Save GAT encoder $Q$
\end{algorithm}

\subsection{GAT Auto-Encoder}
A key reason for the difficulty of DAG scheduling is the dependency and parallelism of graph structure, as we introduced above. An outstanding scheduler needs to maximize parallelism as much as possible under dependency constraints. Thus, the first step in designing a scheduler is to fully grasp the characteristics of the graph structure. Unlike images or texts that can be represented as a regular form in Euclidean space (matrixes or vectors), graph-structured data doesn't have translation invariance, which makes Convolutional Neural Network (CNN) or Recurrent Neural Network (RNN) fails to work well on them.

Graph Neural Networks (GNNs)\cite{GCN,GAT,graphsage,vgae,dgi} have show their abilities in graph-based tasks like node classification, edge classification, edge prediction and graph classification. There are many types of GNN like GCN\cite{GCN}, GAT\cite{GAT}, GraphSAGE\cite{graphsage}, etc. In this work, we choose GAT as our GNN for the following reasons.
Our primary desire is based on its well-behaved performance in widespread graph-based tasks and simple network structure. By introducing attention mechanism, GAT can aggregate the features of node's neighbors to update its own features along the adjacency matrix which implies the graph structure information. Meanwhile, the operation is efficient, since it is parallelizable across node-neighbor pairs. Another advantage of GAT is that it is directly applicable to \emph{inductive} learning problems\cite{GAT}, including tasks where the model has to generalize to completely unseen graphs. 

Another question is how to utilize GAT with DRL in MEC.
Coupling various neural networks into DRL's policy network or value network is a straightforward thought \cite{dagTC2021,dagIOT2022,lan2022generalization}. However, it may contradict the mainstream view that DRL's network should try to be simple, otherwise, the training of it will be unstable and hard to convergence \cite{DQN,DDPG,TD3,PPO,SAC}. Besides, the number of parameters of GNN is generally more than common neural networks (like Multilayer Perceptron, MLP) that consists of several linear layers, which makes the training of it require more computation resources.

A promising approach is to pretrain a GAT encoder in the cloud layer through unsupervised learning. The GAT encoder is used to capture the graph features and translate graph profiles to State vectors that DRL can exploit further. 
By collecting data from edges, the cloud can generate a massive and widely distributed graph datasets. Besides, the datasets can be expanded iteratively as needed. To facilitate the description of the methodology, here we do not consider the quality and reliability of the data, since there are lots of works focus on these issues. Thus, the problem of unsupervised learning requiring large amount of data can be solved properly in MEC. Note that the GAT encoder is to get a comprehensive State representation, so the training of it is totally decoupled from that of DRL. Once the cloud has trained a new GAT encoder, it sends the new GAT encoder down to the edge to replace the old one. This process doesn't affect the ongoing scheduling, but is helpful for the next scheduling, that is, the GAT encoder is incentive for the scheduler.
Next, we will illustrate the implementation of GAT Auto-Encoder.

GAT consists of several stacked graph attention layers. The input of layer is a set of node features, $\bm{h} = \left \{ \vec{h}_1, \vec{h}_2, ..., \vec{h}_{N} \right \}, \vec{h}_i \in \mathbb{R}^F$, where $N$ is the number of nodes of $G$, $\vec{h}_i$ is the feature vector of node $\nu_i$, and $F$ is the number of features in each node. The layer produces a new set of node features (of potentially different cardinality $F^{(l)}$ in layer $l$), $\bm{h^{(l)}} = \left \{ \vec{h}_1^{(l)}, \vec{h}_2^{(l)}, ..., \vec{h}_{N}^{(l)} \right \}, \vec{h}_i^{(l)} \in \mathbb{R}^{F^{(l)}}$ as its output and input of next layer.
The goal of graph attention layers is to aggregate features of neighbors of node to its own features and project node's features into a high-level representation.
Thus, a learnable weight matrix $\bm{W}\in \mathbb{R}^{F^{(l)} \times F}$ is applied to every node and a shared attentional mechanism $a: {\mathbb{R}^{F^{(l)}}} \times {\mathbb{R}^{F^{(l)}}} \rightarrow \mathbb{R}$ is performed to compute \emph{attention coefficients}
\begin{equation}
  e_{i,j}^{(l)} = a(\bm{W}\vec{h}_i, \bm{W}\vec{h}_j)
\end{equation}
that indicates the \emph{importance} of $\nu_j$ to $\nu_i$, where \emph{l} is the index of layer.
Graph structure is injected into the mechanism by performing \emph{masked attention} with the help of the adjacent matrix $adj$.
The adjacent matrix is a $N \times N$ matrix with elements 0 or 1, where the element $adj_{i,j}$ in row $i$ and column $j$ equal to 1 indicates a directed edge from $\nu_i$ to $\nu_j$, and vice versa.  Thus, it is an asymmetric matrix with all diagonal elements 1.
Following \cite{GAT}, we only compute $e_{i,j}$ for the first-order neighbors of $\nu_i$, that is, $adj_{j,i}=1$.
After going thorough all $L$ layers, $G$'s node representation $\bm{\hat{h}} = \bm{h^{(L)}} = \left \{ \vec{\hat{h}}_1, \vec{\hat{h}}_2, ..., \vec{\hat{h}}_{N} \right \}$ is obtained. 
\begin{figure*}[t] 
  \centering
  \includegraphics[width=0.92\textwidth]{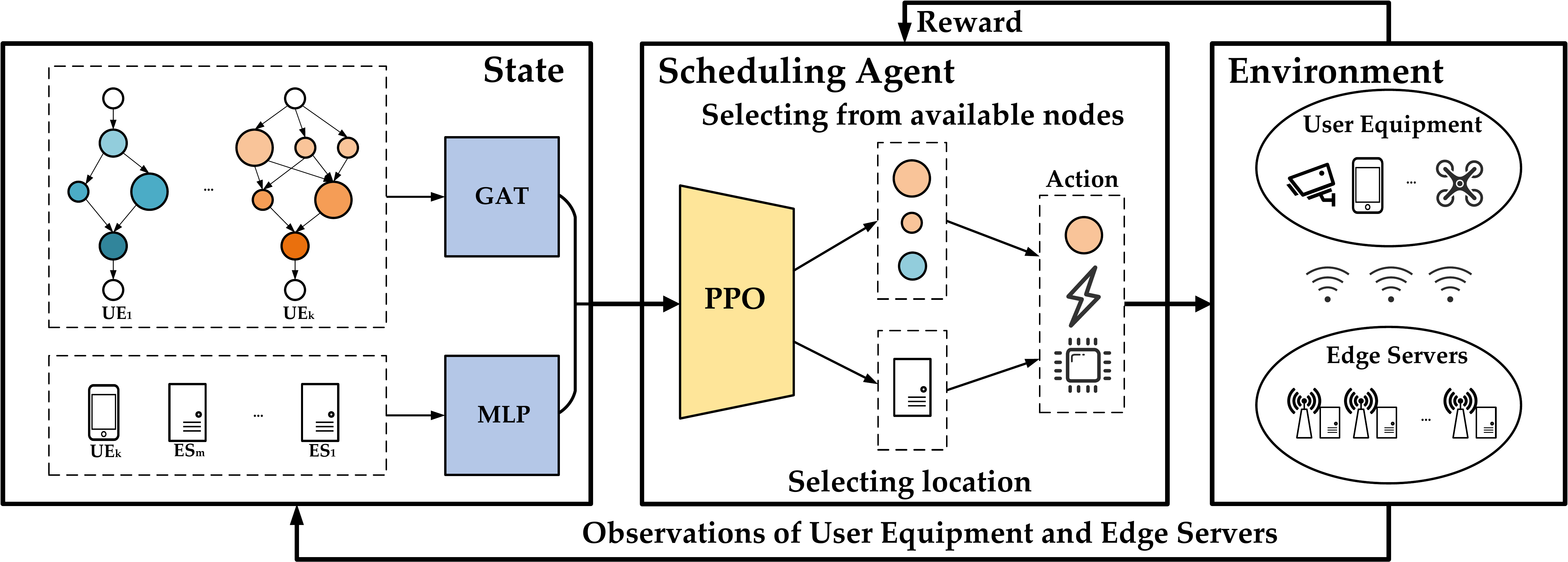}
  \caption{The framework of the DTODRL. A multi-user multi-edge scenario is considered here. The size of node represents its computation density. The larger the node, the greater the computation density, and vice versa. Observations of the environment are needed to be represented as input vectors firstly for the policy network, which then outputs actions.}
  \label{fig:structure}
\end{figure*}

Note that the GAT encoder is used to get DAG's State representation, therefore, we need to evaluate whether the output representation is good enough, i.e., whether it is distinguished. A basic idea is that similar input node features may have similar representations in the new vector space as well, and the representations of adjacent nodes may have higher similarity compared to that of non-adjacent nodes due to the attention mechanism. Thus, we utilize a decoder with similar structure to the encoder after \cite{gate}, which reverses the encoding process to reconstruct node attributes. Correspondingly, $\bm{\hat{h}}$ is the input of the decoder, $\bm{\hat{h}^{(l)}}$ is the output of decoder's $l$th layer, and $\bm{h^{\prime}} = \bm{\hat{h}^{(L)}} = \left \{ \vec{h}_1^{\prime}, \vec{h}_2^{\prime}, ..., \vec{h}_{N}^{\prime} \right \}$ is the final output of the decoder.
Based on the above idea, the loss function that consists of two parts: \emph{Feature Loss} and \emph{Structure Loss} is adopted to optimize both encoder and decoder.
Feature Loss implies the consistency of reconstructed node features and original node features, which is represented as:
\begin{equation}
  \label{eq:featureloss}
  Feature Loss = \sum_{i=1}^N \Vert \vec{h}_i - \vec{h}_i^{\prime} \Vert_2.
\end{equation}
Structure Loss aims to make the representations of neighboring nodes similar as follows:
\begin{equation}
  \label{eq:structureloss}
  Structure Loss = -\sum_{i=1}^N\sum_{j\in \mathcal{N}_i} log \left(\frac{1}{1+exp(-\vec{\hat{h}}_i^T \vec{\hat{h}}_j)} \right),
\end{equation}
where $\mathcal{N}_i$ is the first-order neighbors of $\nu_i$, $T$ is transpose operation.
Network parameter optimization is conducted both on the encoder and the decoder, but only the encoder is kept.
The main process of GAT auto-encoder's training is presented in Algorithm \ref{algo:encoder}. This unsupervised learning style normally requires numerous data and computation resources\cite{bert,mae}, so we put the training of it on the resourceful cloud which can collect a lot of data from multiple edges. In additional, this pretrained encoder can continuously optimize itself with the increase of collected data and are sent down to edge layer periodically, so its iteration won't disrupt the scheduling between edge layer and user layer.


\subsection{DRL Design}
In this work, we consider a multi-user multi-edge scenario where each UE can communicate with all ESs, and ESs can communicate with each other. 
Fig. \ref{fig:structure} gives the structure of the DTODRL, the DTODRL consists of three main parts: \emph{Environment}, \emph{State} and \emph{Scheduler}.
Raw environment observations are encoded to get State representation firstly. Then, State is sent to Scheduler as input to get an Action. Each Action will select a legal (unscheduled and all of its predecessor nodes scheduled) node, and meanwhile its corresponding offloading location.
After conducting the Action with Environment, it will produce a feedback reward to guide Scheduler updated better.
By adapting DRL to solve DAG task offloading problem,
we need to model it as a Markov Decision Process (MDP).
Next, we will give the state space,
action space, reward function of MDP as follows.

\subsubsection{State Space}
State is regarded as the input of DRL. Normally, it can display the situation of the system model, guide DRL agent to do action and is represented by attributes of the system model. But in this work, the system model contains a graph structure task which has node-level attributes and graph structure attributes, making it hard to abstract into regularized data. The GAT encoder as we introduced above works here. It can fuse graph structure attributes into node-level attributes, so we can use a set of attributes of nodes to represent overview task state.
For the sake of distinction, we name the raw attributes of the system model Observation and the final input of DRL State. Both of them exhibit the system model, but State is in a deeper feature space. The system model is composed of two main parts: task and execution location.
Both DAG of task and available offloading locations are need to be taken into account when scheduling.
We firstly define the observation of a task:
\begin{equation}
  o_{task} = \left\{o_{node}, adj \right\},
\end{equation}
where $adj$ is the adjacency matrix of task and $o_{node}$ is a set of profiles of nodes. Profile vector of node $\nu_i$ consists of six features: 
1) $C_i$; 2) $D_i$; 3) in\_degree: how many directed edges to $\nu_i$; 4) out\_degree: how many directed edges from $\nu_i$ ;
5) $loc_i$: the location of $\nu_i$ if it's scheduled; 6) $ava_i$: if $\nu_i$ is available to schedule.
Among them, $C_i$, $D_i$, in\_degree and out\_degree remain their initial values, while $loc_i$ and $ava_i$ will change from initial value -1 (means that $\nu_i$ hasn't been scheduled) and 0 (means that $\nu_i$ has been scheduled or some of its predecessor nodes unscheduled)
as scheduling proceeds. In the beginning of scheduling, $ava$ attribute of successor nodes of \emph{Start} node are set to 1, which means they are prepared to be scheduled.
Once an action that select $\nu_i$ has interacted with the environment, $loc_i$ will change to the corresponding location index and $ava$ attribute of a new set of nodes updates to 1 depend on previous scheduling. Then, the observation of available offloading locations is defined as their resource-related profiles. Profile vector of an available offloading location (UE or ES) consists of two features: 1) $EFT$: the estimated free time of it; 2) $f$: the computation power of it. 
The observation of all available offloading locations is represented as:
\begin{equation}
  o_{locations} = \left\{o_{location}| o_{location}=(EFT, f)\right\}.
\end{equation}
Next, we need to convert raw observations to state representations.
Correspondingly, the state space is defined as a combination of two parts: 
\begin{equation}
    S=\left\{ s_t|s_t=\left( s_{task}, s_{locations} \right) \right\}, 
\end{equation}
\begin{equation}
  s_{task} = GAT( o_{task}),
\end{equation}
\begin{equation}
  s_{locations} = MLP( o_{locations}),
\end{equation}
where $s_{task}$ is the state of DAG and $s_{locations}$ is the state of available offloading locations for all. 
$s_{task}$ and $s_{location}$ are outputs of GAT and MLP respectively.
In order to fusion $s_{node}$ and $s_{location}$ together, a \emph{flatten} layer are put in the end of GAT, then, we concatenate them as the final state representation in each time step.

\begin{figure*}[t] 
  \begin{center}
    \includegraphics[width=0.95\textwidth]{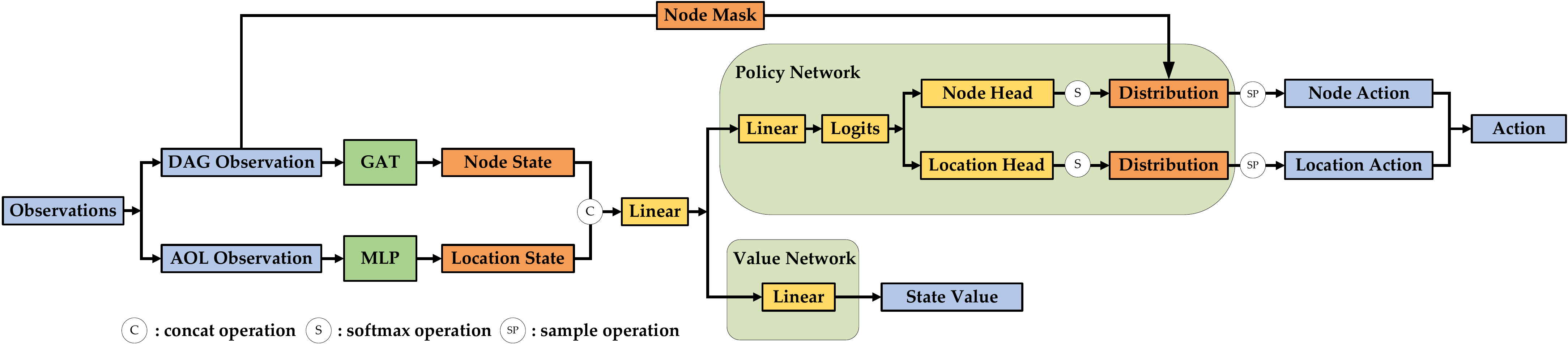}
  \end{center}
  
  \caption{The neural network view on how the DTODRL represents state and generates action.}
  \label{fig:algo}
\end{figure*}

\subsubsection{Action Space}
In each time step of scheduling, DTODRL will select a node and its corresponding offloading location. They form a complete action together, so the action space consists of two parts:
\begin{equation}
    A=\left\{ a_t|a_t=\left( a_{node}, a_{location} \right) \right\},
\end{equation}
where $a_{node}$ is the index of selected node, and $a_{location}$ is its corresponding offloading location.
We use two different output heads named \emph{Node Head}  and \emph{Location Head} which share a same state representation.
To ensure selecting nodes does not break dependencies among nodes, a node mask vector is used to filter unavailable nodes (at least one of their predecessors unscheduled) and keep output dimension (number of nodes $N$) of \emph{Node Head} unchanged in each time step. When an action interacts with the environment, a new node mask vector and state are updated in the same time.
For each node in $ue_k$, its available offloading location set is $AOL_k=\left\{ ue_k,es_1\,\,,...,es_M \right\} $. To keep the dimension of \emph{Location Head} from increasing with the number of users, we use dimension 0 to represent $ue_k$ itself. Thus, the output dimension of \emph{Location Head} is fixed to \emph{M + 1}.
After all actions are generated, a complete scheduling plan is formed.

\subsubsection{Reward Function}
The objective of our scheme is to minimize mean \emph{AFT} of all tasks.
Assume that there is no scheduling, all nodes will be executed locally. We take local execution as a baseline, a
well-behaved scheduling should produce a mean \emph{AFT} as smaller than the baseline as possible. 
As we discussed above, each action will select a single node,
let $A_{1:t}$ denote a sequence of actions which already scheduled $t$ nodes from time step 1 to time step $t$. This sequence is called scheduled part. If scheduler stops here, then rest $N-t$ nodes are executed locally by default which is called local part.
In each time step of the DTODRL, there are scheduled part and its corresponding local part. By combining two parts together, a complete scheduling plan for all nodes are obtained and \emph{Estimated Finish Time (EFT)} of all tasks can be calculated.
Thus, the reward of the DTODRL in each time step can be regarded as the reduction of calculated \emph{EFT}:
\begin{equation}
  r_t = \frac{1}{K} \left(\sum_{k=1}^K {EFT}^{A_{1:t-1}}_k - \sum_{k=1}^K {EFT}^{A_{1:t}}_k\right),
\end{equation}
where $A_{1:t-1}$ is a sequence of actions before $t$'s action and $A_{1:t}$ is the sequence containing $t$'s action, $K$ is the number of users.

\begin{algorithm}
  \caption{The Training Process of PPO-based Algorithm}
  \label{algo:ppo}
  \textbf{Initialize} the old policy $\pi_{\theta_{old}}$, the current policy $\pi_{\theta}$
  $\theta_{old} \leftarrow \theta$ \\
  \For{$i=1,2,...$}{
    Collect the set of trajectories $\mathcal{D}_i$ by running policy $\pi_{\theta_{old}}$ in the environment. \\
    \For{each trajectory $\tau \in \mathcal{D}_k$}{
      Compute target state values $V_{targ}(s_1),..., V_{targ}(s_O)$ according to Eq. \ref{eq:vtarg}.\\
      Compute advantage estimates $\hat{A}_1,...,\hat{A}_O$ according to Eq. \ref{eq:delta} and Eq. \ref{eq:adv}. \\
      Store target state values and advantage estimates in $\mathcal{D}_i$.
    }
    \For{$k=1,2,...$}{
      Sample minibatches from $\mathcal{D}_i$. \\ 
      Compute the policy loss according to Eq. \ref{eq:policy}\\
      Compute the value function loss according to Eq. \ref{eq:vf}\\
      Optimize the target function $L^{PPO}$ w.r.t. $\theta_k$ by taking minibatch stochastic gradient descent. \\
    }
    $\theta_{old} \leftarrow \theta$ \\
  }
  
\end{algorithm}

\subsubsection{DRL Training}
The goal of training a DRL algorithm is to exploit a near-optimal policy that provides the maximal accumulated reward.
In this work, we select PPO \cite{PPO} as our based DRL algorithm. As the most promising and widely used DRL algorithm that utilizes generalized advantage estimation (GAE) \cite{gae} and clipped surrogate function \cite{PPO}, PPO is easy to train, not sensitive to hyperparameters and can be applied to multi-dimensional action space. Next, we will illustrate how PPO works and its training process.

PPO belongs to policy-based DRL that normally aims to optimize the policy network by computing an estimator of the \emph{policy gradient (PG)} and plugging it into a gradient ascent algorithm. 
To improve the utilization efficiency of sampling trajectories, PPO
generates trajectories using the old policy $\pi_{\theta_{old}}$ and updates the current policy $\pi_{\theta}$ from the initial values as $\pi_{\theta_{old}}$, then synchronize them after several episodes. To make sure the trajectories generated by $\pi_{\theta_{old}}$ can be used to update $\pi_{\theta}$, PPO uses an importance sampling estimator to compensate for the gap between the training data distribution and the current policy state distribution, which is the probability ratio term:
\begin{equation}
  {pr}_t(\theta) = \frac{\pi_{\theta}\left( a_t|s_t\right)}{\pi_{\theta_{old}}\left( a_t|s_t\right)},
\end{equation}
so, ${pr}_t(\theta_{old})=1$.
Furthermore, in order to avoid a large update of the policy, PPO applies a clipped surrogate function to limit the range of the optimization target that is given by:
\begin{small}
  \begin{equation}
    \label{eq:policy}
    L^{C}(\theta) = \mathbb{E}\left[\sum_{t=1}^{O}\min {pr}_t (\theta) \hat{A}_t,clip\left({pr}_t (\theta),1-\epsilon,1+\epsilon \right)\hat{A}_t \right],
  \end{equation}
\end{small}
where $\hat{A}_t$ is the estimator of the advantage function at time step $t$ and epsilon is a hyperparameter which indicates how far away the new policy is allowed to go from the old.
There are two terms in the min: unclipped objective ${pr}_t \left( \theta \right) \hat{A}_t$ and clipped objective $clip(\cdot) \hat{A}_t$. $clip(\cdot)$ clips the probability ratio $r_t \left(\theta\right)$ at $1+\epsilon$ or $1-\epsilon$ depending on whether the advantage is positive or negative. By doing so, the final objective is a lower bound on the unclipped objective. 
The advantage for action-state pair depicts whether the action is better than baseline in the state or worse. 
Suppose the advantage is positive, the objective will increase if the action becomes \emph{more} likely---that is, if $\pi_\theta(a|s)$ increases. Correspondingly, Suppose the advantage is negative, the objective will increase if the action becomes \emph{less} likely---that is, if $\pi_\theta(a|s)$ decreases. Thus, maximizing 
$L^C$ results in the rising of the probability of high-advantage actions and the failing of that of low-advantage actions.

In PPO, $\hat{A}_t$ is obtained by utilizing the GAE that is expressed as 
\begin{equation}
  \label{eq:adv}
  \hat{A}_t = \sum _{k=0} ^{O-t+1} (\gamma\lambda)^k(\delta_{t+k}),
\end{equation}
\begin{equation}
  \label{eq:delta}
  \delta_t = r_t + \gamma V_\pi(s_{t+1}) - V_\pi(s_{t}), 
\end{equation}
where $\lambda$ is used to control the trade-off between bias and variance, $\delta_t$ is the TD-error term at time step $t$, $V_\pi(s)$ is the state-value function.
In order to compute variance-reduced advantage-function estimators, PPO applies a learned state-value function $V(s)$ that is a neural network. Besides, an entropy bonus is considered to ensure sufficient exploration. In our implementation, the value function and the policy shares parameters of networks \cite{PPO,improved_PPO}, therefore, the final optimization target is obtained by combing these terms:
\begin{equation}
  L^{PPO} = L^C -c_1 L^{VF} + c_2 S[\pi_\theta](s_t),
\end{equation}
where $c_1$, $c_2$ is coefficient, $S$ denotes an entropy bonus, $L^{VF}$ is a squared-error between predicted state value $V_\pi(s_t)$ and the sum of discounted rewards $\hat{R}_t $.
\begin{equation}
  \label{eq:vf}
  L^{VF}(\theta) = \mathbb{E}\left[\sum_{t=1}^O (V_\pi(s_{t}) - \hat{R}_t)^2 \right],
\end{equation}
\begin{equation}
  \label{eq:vtarg}
  \hat{R}_t = \sum _{k=0} ^{O-t+1} \gamma^k r_{t+k},
\end{equation}
where $\lambda$ is discounted factor. The training process is presented in Algorithm \ref{algo:ppo}. We keep the most of the native PPO unchanged but separate the logits output of the policy into two-dimension discrete action, which won't affect the gradient flows of neural network. Fig. \ref{fig:algo} shows the pipeline of the DTODRL in neural network view.



\begin{table}[htbp]
  \caption{Neural Networks and Training Hyperparameters.}
  \begin{center}
    \resizebox{\columnwidth}{!}{
    \begin{tabular}{c|c|c|c|c}
      \hline
      \textbf{Hyperparameter} & \textbf{Value} & &\textbf{Hyperparameter} & \textbf{Value}\\  
      \hline
      GAT Layers & 3 & & GAT Activation Function & LeakyReLU\\
      \hline
      GAT Hidden Units & 128 & & GAT Learning Rate & $1 \times 10^{-3}$\\
      \hline
      GAT Attention Heads & 3 & &  MLP Hidden Units & 256\\
      \hline
      PPO Loss Coefficient $c_1$ & 0.5 & & PPO Entropy Coefficient $c_2$ & 0.01 \\
      \hline
      Discount Factor $\gamma$ & 0.99 & & Clip Constant $\epsilon$ & 0.2\\
      \hline
      Optimization Method & Adam & & PPO Learning Rate & $3 \times 10^{-4}$\\
      \hline 
      Batch Size & 512 & & PPO Activation Function & Tanh\\
      \hline

    \end{tabular}
    }
  \end{center}
  \label{table:network}
\end{table}

\section{Experiments}
This section presents experiment results and performance analysis. We firstly illustrate how the simulation data is generated, then give the results of our proposed scheme the \emph{DTODRL} and other methods in comparison of different experiment settings. Finally, we analyze the results.

\begin{table*}
  \centering
  \caption{The comparisons of the \emph{DTODRL} and other algorithm for different \emph{node\_number} (from 10 to 50 with a step of 5) under single-user single-edge scenario.}
  \begin{tabular}[b]{cccccccc}
  \toprule[1pt]     
  \# of nodes & Optimal & Local & Remote & RR & Random & HEFT-based & DTODRL \\ 
  \midrule[.5pt]  
  
  10 & \textbf{275.5} & 529.06 & 552.53 & 446.58 & 470.06 & 379.97 & \textbf{284.48}\\
  15 & \textbf{366.8} & 784.98 & 781.28 & 571.45 & 620.95 & 519.39 & \textbf{397.56}\\
  20 & N/A & 1068.03 & 975.77 & 729.19 & 820.69 & 627.13 & \textbf{518.51}\\
  25 & N/A & 1310.64 & 1210.26 & 864.28 & 981.67 & 739.05 & \textbf{656.49}\\
  30 & N/A & 1604.89 & 1403.18 & 1032.8 & 1086.92 & 873.63 & \textbf{769.61}\\
  35 & N/A & 1889.81 & 1605.74 & 1122.46 & 1231.42 & 1043.3 & \textbf{885.21}\\
  40 & N/A & 2132.53 & 1842.79 & 1322.46 & 1534.84 & 1192.19 & \textbf{1090.72}\\
  45 & N/A & 2417.51 & 2036.82 & 1441.7 & 1660.29 & 1266.46 & \textbf{1194.86}\\
  50 & N/A & 2678.94 & 2266.63 & 1639.24 & 1949.14 & 1466.76 & \textbf{1352.3}\\
  \bottomrule[1pt]    
  \label{table:result}
\end{tabular}
\end{table*}

\begin{figure}[htbp]
  \vspace{-0.5cm}
  \centering
  \begin{subfigure}{0.9\linewidth}
    \centering
    \includegraphics[width=0.9\textwidth]{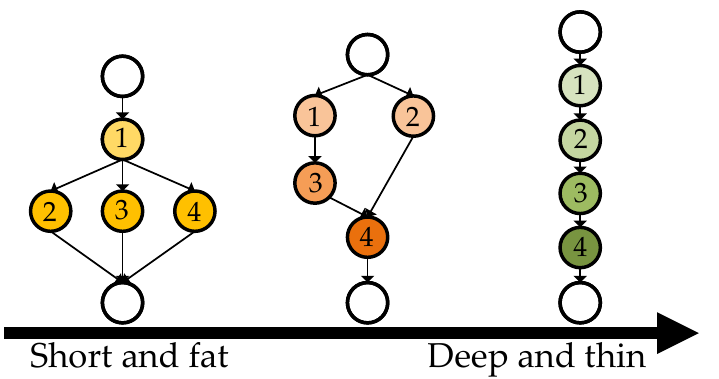} 
    \caption{Example of how the shape of DAGs with 4 nodes varies.}
  \end{subfigure}

  \begin{subfigure}{0.9\linewidth}
    \centering
    \includegraphics[width=0.9\textwidth]{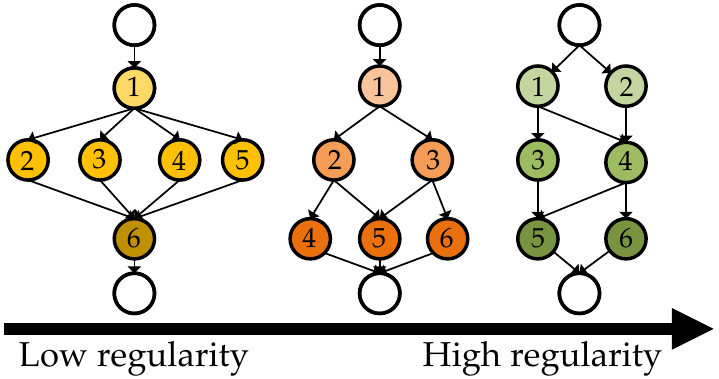}
    \caption{Example of how the regularity of DAGs with 6 nodes varies.}
  \end{subfigure}
  
  \caption{Different DAGs generated by the generator with 4 parameters.}
  \label{fig:dag}
\end{figure}

\begin{figure*}[ht]
  \centering
  
  \begin{subfigure}{0.49\linewidth}
    \centering
    \includegraphics[width=0.9\linewidth]{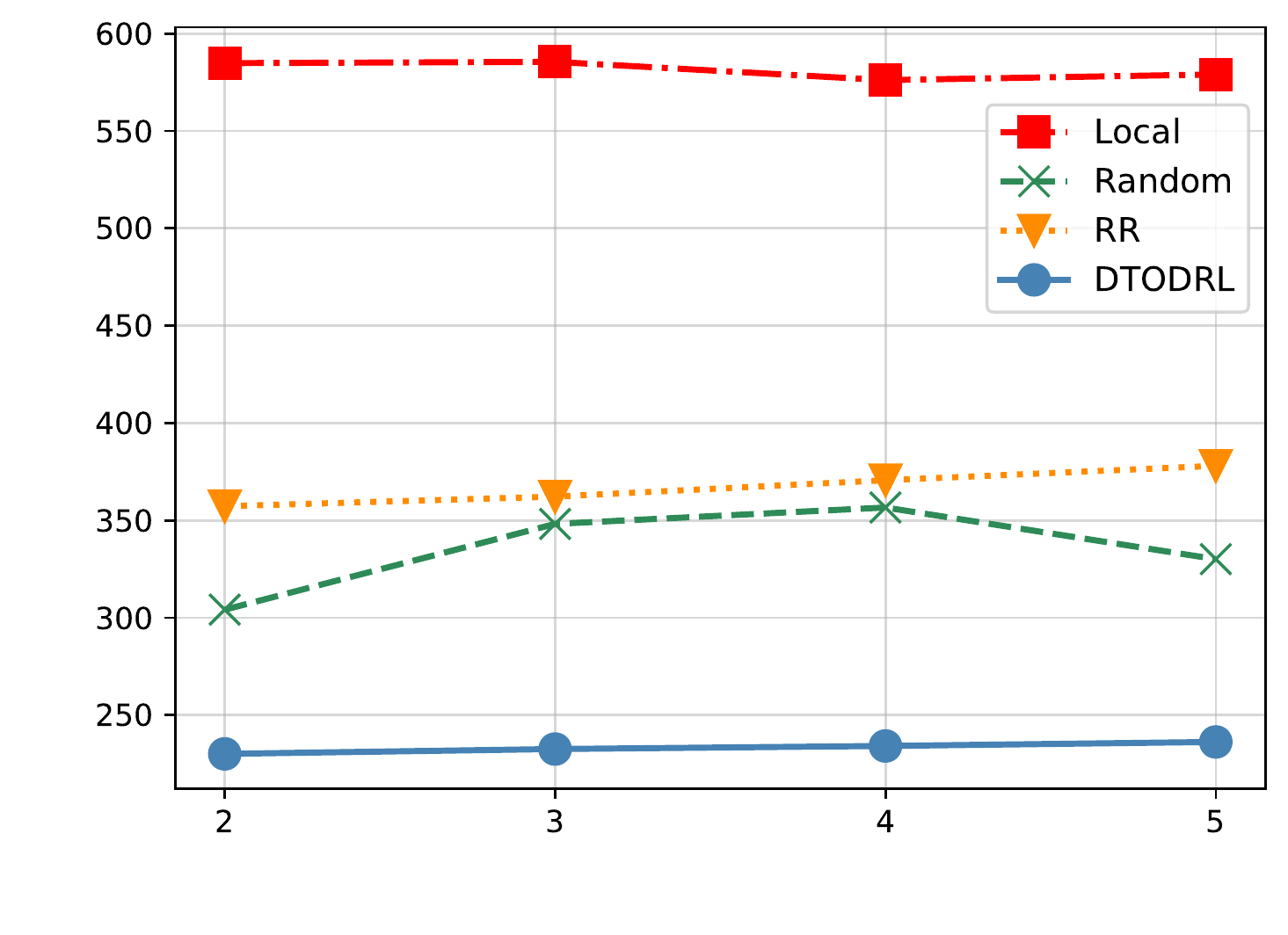} 
    \caption{10 nodes for each UE's DAG.}
    \label{subfig:3-1}
  \end{subfigure}
  \begin{subfigure}{0.49\linewidth}
    \centering
    \includegraphics[width=0.9\linewidth]{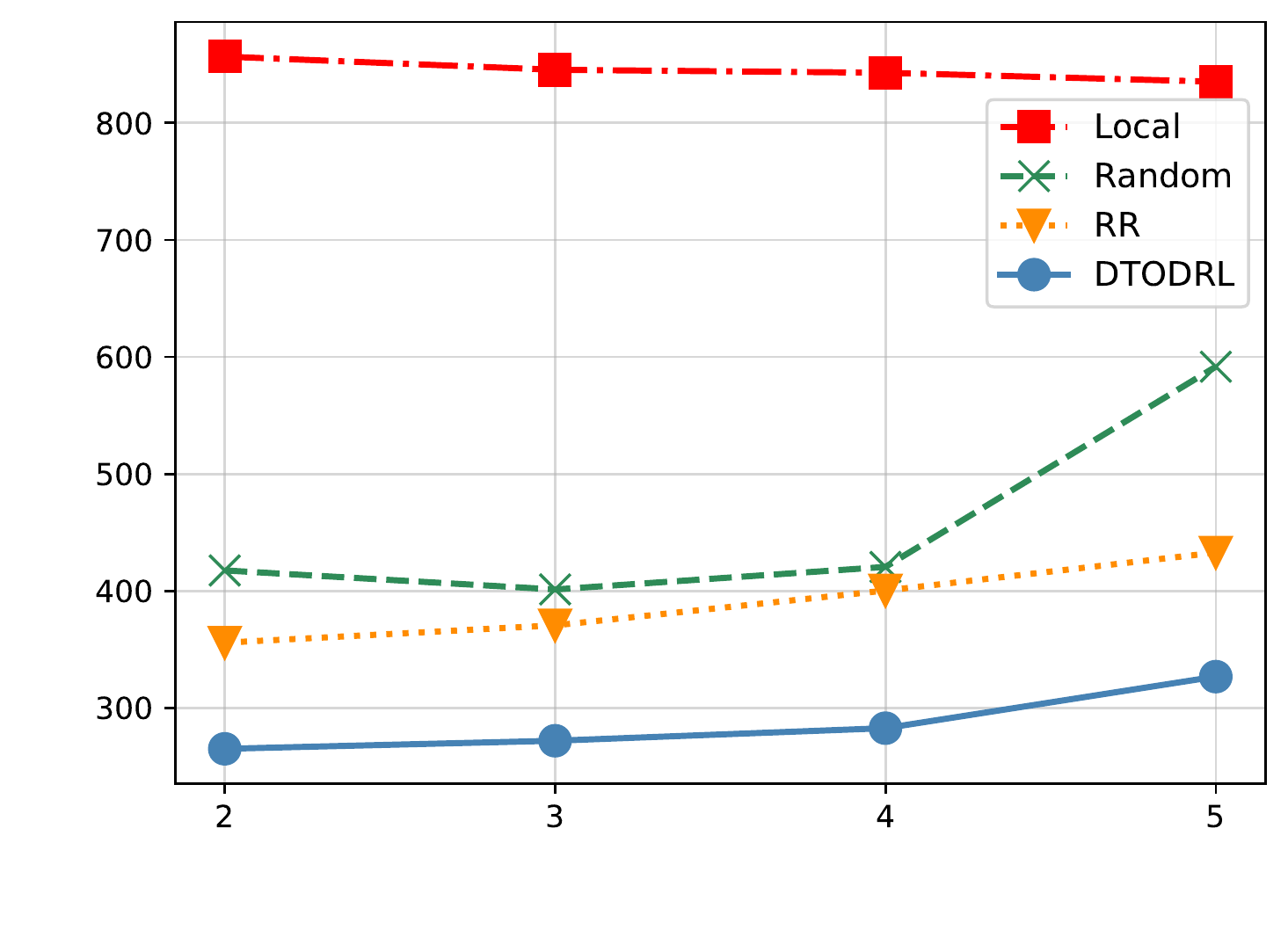}
    \caption{15 nodes for each UE's DAG.}
    \label{subfig:3-2}
  \end{subfigure}
  \begin{subfigure}{0.49\linewidth}
    \centering
    \includegraphics[width=0.9\linewidth]{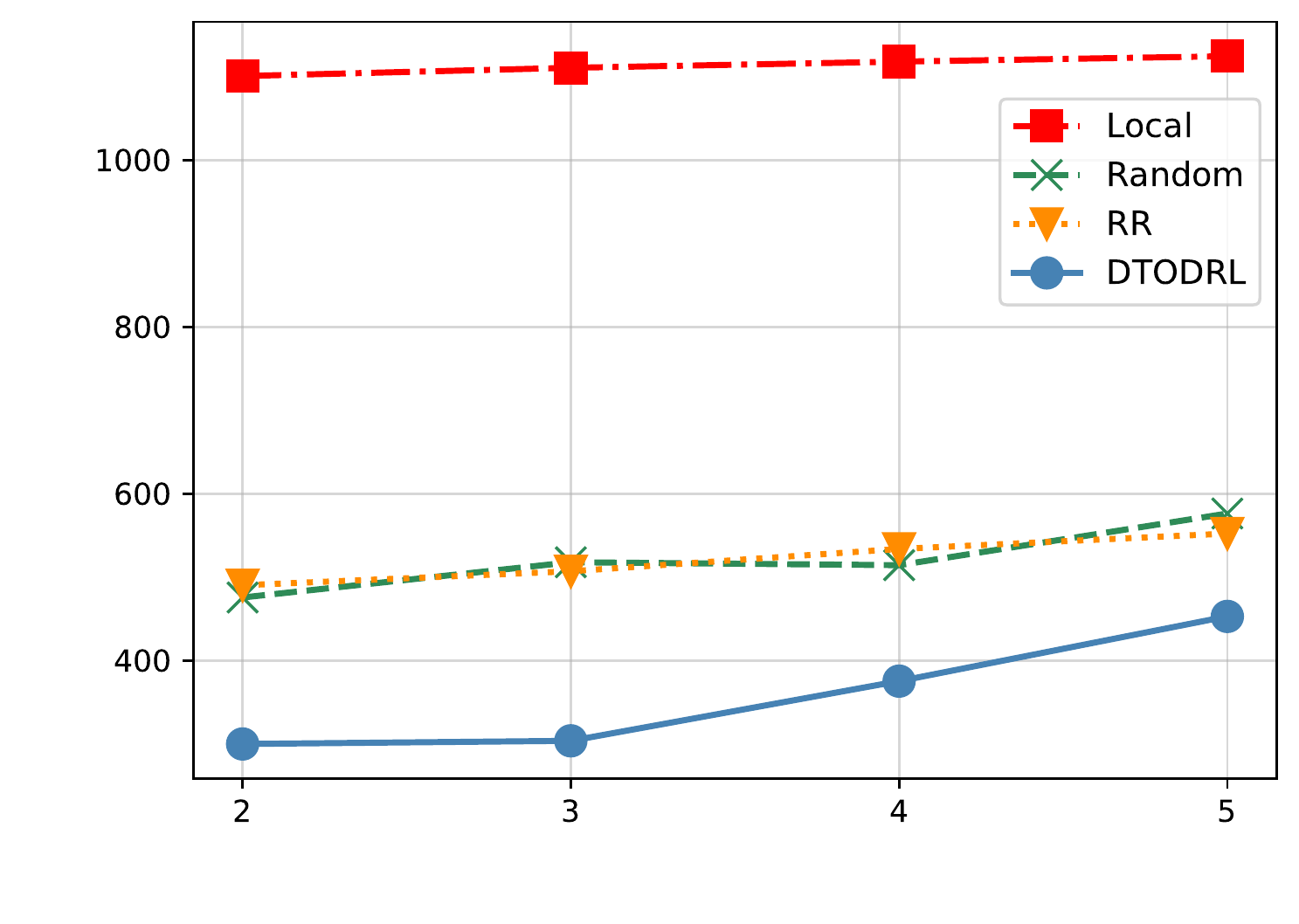}
    \caption{20 nodes for each UE's DAG.}
    \label{subfig:3-3}
  \end{subfigure}
  \begin{subfigure}{0.49\linewidth}
    \centering
    \includegraphics[width=0.9\linewidth]{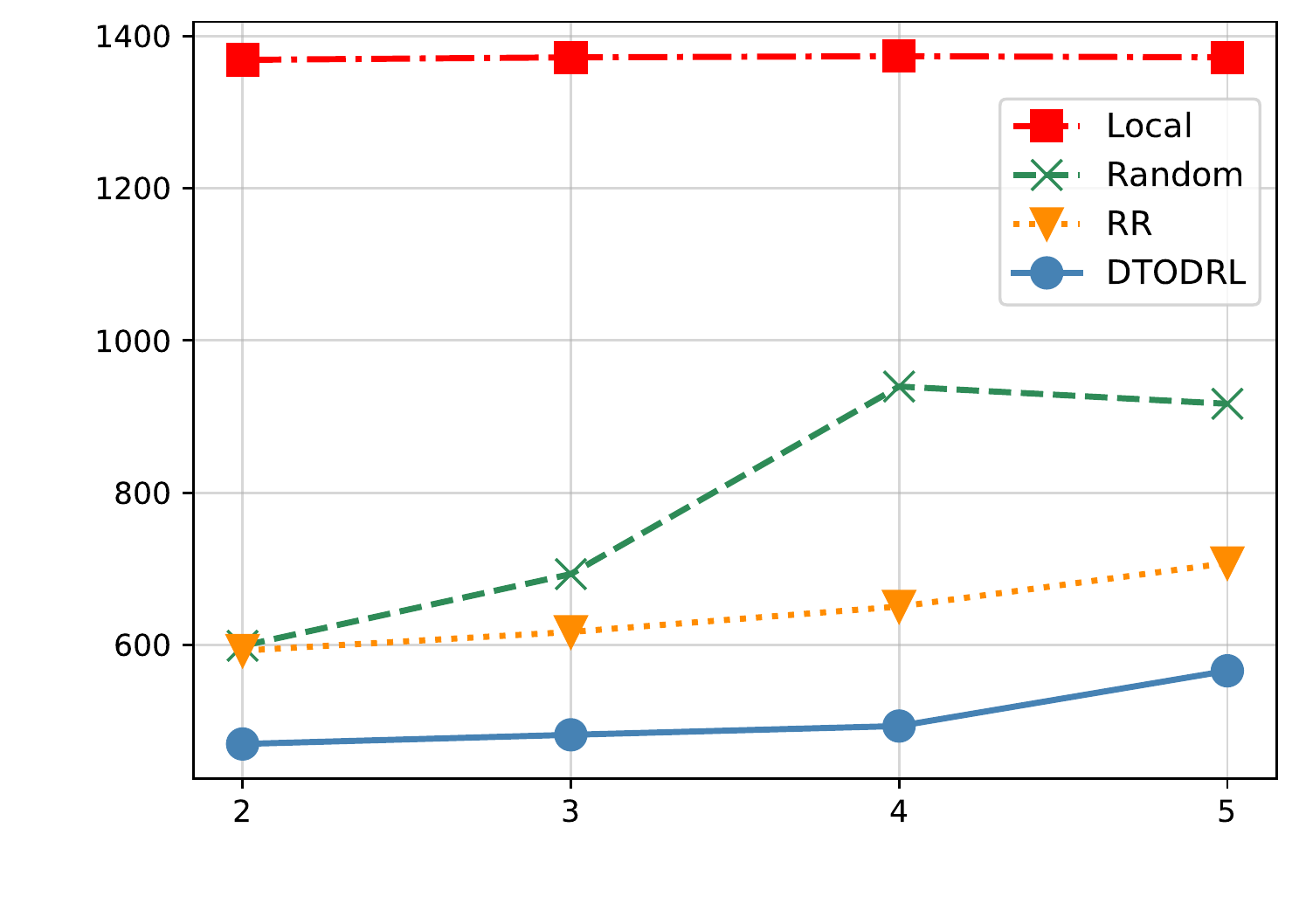}
    \caption{25 nodes for each UE's DAG.}
    \label{subfig:3-4}
  \end{subfigure}
  
  \caption{The comparisons of the DTODRL and other algorithms in terms of mean \emph{AFT} of 2 $\sim$ 5 UEs with 10, 15, 20 and 25 nodes for each DAG when the number of ESs equal to 3.}
  \label{fig:diffenrt_number_of_nodes_and_servers}
\end{figure*}

\subsection{Simulation Setting}
\label{sec:dag_generator}
In our simulation, we consider both single-user single-edge and multi-user multi-edge scenario. 
We use DAG to represent UE's task, where each node represents the subtask of it, and each directed edge represents the dependency between subtasks. To ensure that simulation is wide enough
to prove the applicability of our method, we use the DAG generator described in \cite{HEFT,PEFT} to generate various graph structures. There are four parameters to control the generated graph structure:
\emph{node\_num}, \emph{max\_out\_degree}, \emph{$\alpha$} and \emph{$\beta$}.
\emph{node\_num} determines the total number of nodes, and \emph{max\_out\_degree} limits the max out degree of nodes. \emph{$\alpha$} named \emph{shape} parameter is used to control the depth of DAG and number of nodes of each level. \emph{$\beta$} named \emph{regularity} parameter determines the uniformity of the number of tasks in each level.
By setting the range of four parameters, different forms of DAG can be generated. Fig. \ref{fig:dag} gives two examples of different shapes of DAG, given the number of nodes.

Besides graph structure, nodes and edges have their own attributes. We set the required CPU size $C_i$ of $\nu_i$ to random value of $10^8 \sim 10^9$ cycles per sec and required data size $D_i$ of it to $20 \sim 200$ KB. $D_{j,i}$ is the attribute of edge and is set between $100 \sim 500 $ KB.
After generating the graph, we assign random attribute values to its nodes.
For multi-user setting, we firstly generate $K$ separate DAG, then merge them by merging their private \emph{Start} node, which ensures that the hierarchical position and dependency of each node on the original graph and the merge graph remain unchanged. 
All code of experiments are implemented via PyTorch\cite{} and dgl (a GNN library) \cite{}, Table \ref{table:network} gives the value of the neural network and training hyperparameters. Most of these parameters remain the same as the general parameter settings used by PPO and GAT for other tasks\cite{GAT,PPO,improved_PPO}.

\begin{figure*}[ht]
  \centering
  
  \begin{subfigure}{0.49\linewidth}
    \centering
    \includegraphics[width=0.9\linewidth]{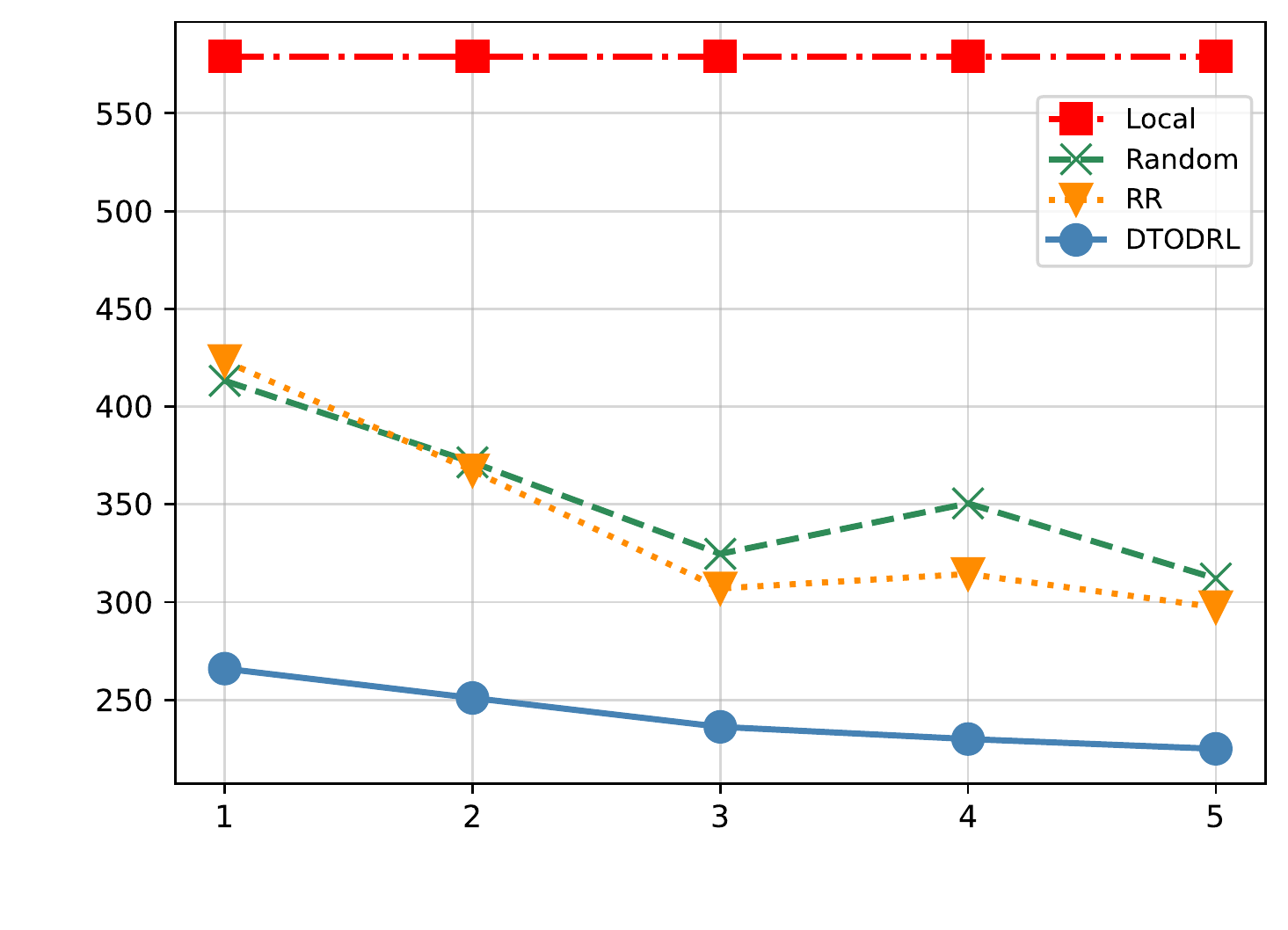} 
    \caption{10 nodes for each UE's DAG.}
    \label{subfig:2-1}
  \end{subfigure}
  \begin{subfigure}{0.49\linewidth}
    \centering
    \includegraphics[width=0.9\linewidth]{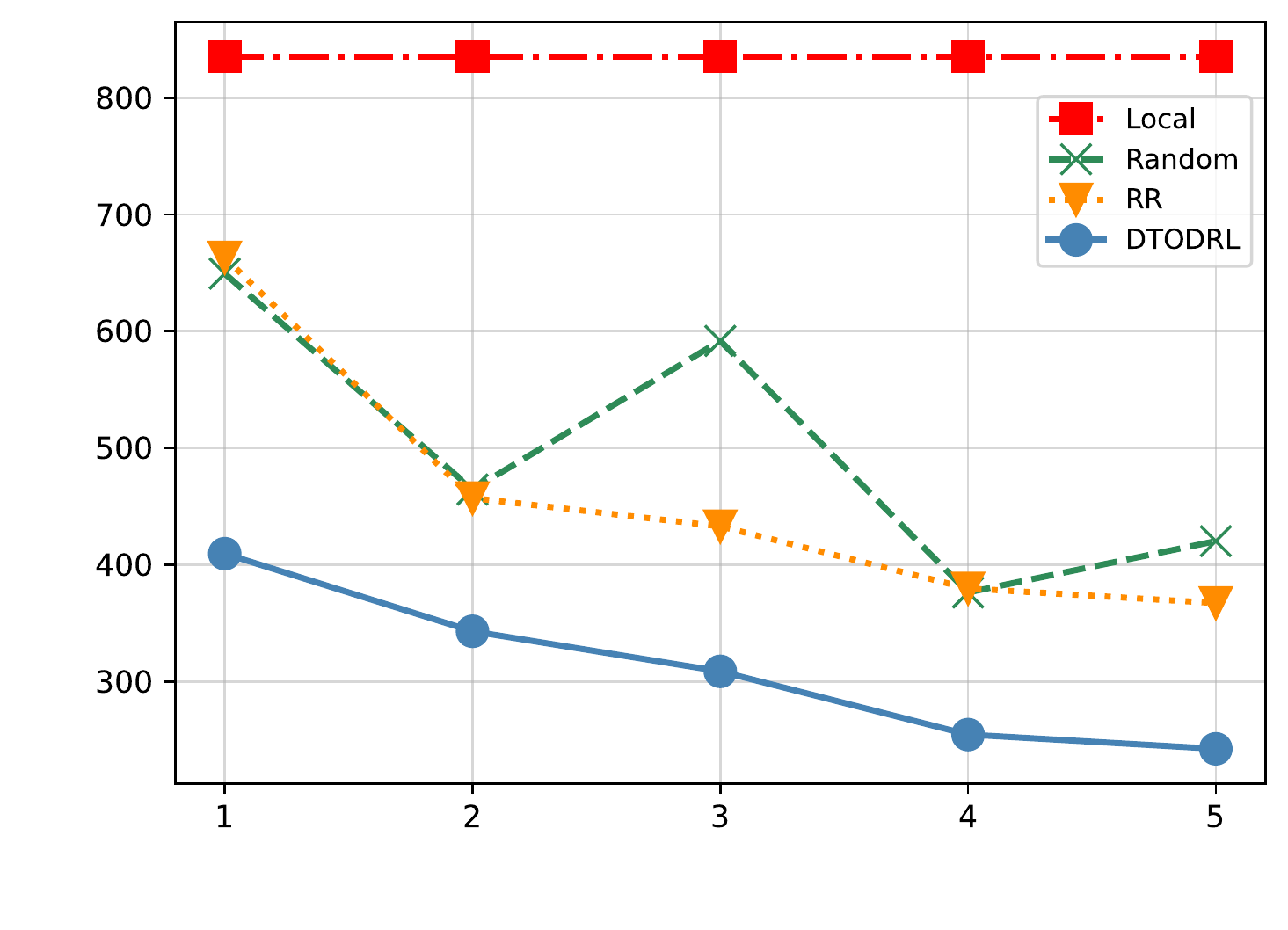}
    \caption{15 nodes for each UE's DAG.}
    \label{subfig:2-2}
  \end{subfigure}
  \begin{subfigure}{0.49\linewidth}
    \centering
    \includegraphics[width=0.9\linewidth]{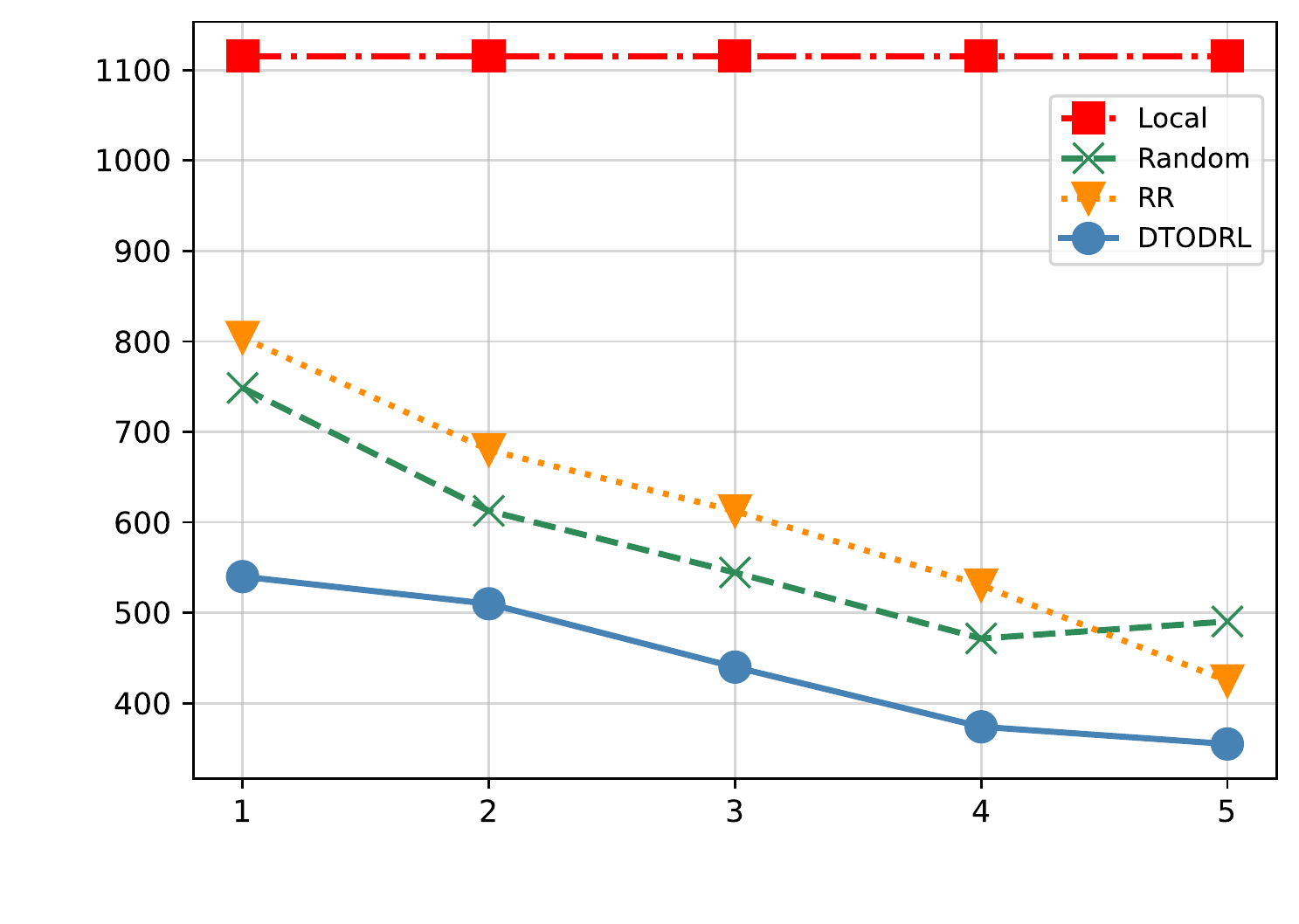}
    \caption{20 nodes for each UE's DAG.}
    \label{subfig:2-3}
  \end{subfigure}
  \begin{subfigure}{0.49\linewidth}
    \centering
    \includegraphics[width=0.9\linewidth]{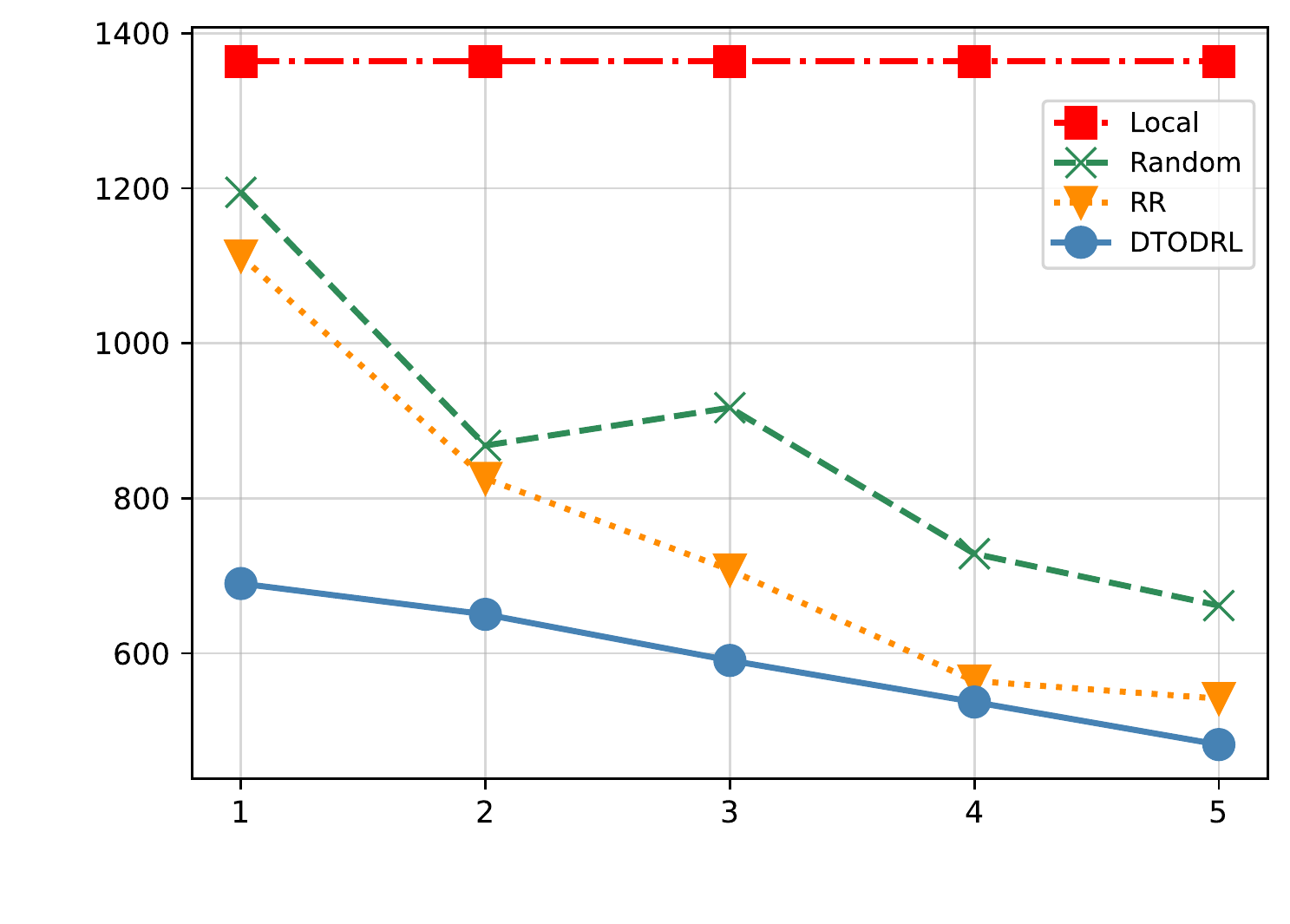}
    \caption{25 nodes for each UE's DAG.}
    \label{subfig:2-4}
  \end{subfigure}
  
  \caption{The comparison of the DTODRL and other algorithms in terms of mean \emph{AFT} of 5 UEs with 10, 15, 20 and 25 nodes for each DAG when the number of ESs range from 1 to 5.}
  \label{fig:diffenrt_number_of_nodes_and_servers}
\end{figure*}

\subsection{Compared Baselines}
As far as we know, we are the first to study in DTO problem under multi-user multi-edge scenario. Three baselines are used in comparison: \emph{Local}, \emph{Round-Robin (RR)} and \emph{Random}. Specially for single-user single-edge scenario, \emph{Remote} is also compared, and it is possible to find the optimal solution by exhaustive search. 
Additional, a heuristic-based method named HEFT\cite{HEFT} under simple scenario are also considered as a competitor. 
These compared methods are as follows:
\begin{itemize}
  \item \emph{Optimal}: Find out the optimal offloading decision for all nodes and their scheduling order in the same time by exhaustive search.
  \item \emph{Local}: All nodes are executed locally in the order of topological sorting index.
  \item \emph{Remote}: All nodes are executed in ES in the order of topological sorting index.
  \item \emph{Round-Robin (RR)}: Nodes are scheduled in index order, and their corresponding locations are also selected in index order.
  \item \emph{Random}: Randomly select node from available nodes and randomly select its corresponding location.
  \item \emph{HEFT}: Nodes are firstly ranked based on Heterogeneous Earliest-Finish-Time (HEFT) \cite{HEFT}. At each step, HEFT selects the task with the highest upward rank value and then select local or edge server that minimize estimated finish time. 
\end{itemize}

\subsection{Evaluation under Single-user Single-edge Setting}
We investigate performances of the \emph{DTODRL} and other 5 methods under single-user single-edge scenario in this section. In this case, we fix $f^{ES}$ of the only ES with one processor to 10Ghz, $f^{UE}$ to 1Ghz, and transmission rate between UE and ES $tr^l$ to 2 Mbps.
Experiments are conducted on 50 random gerated graphs for each \emph{node\_number}, and the results are averaged.

Table. \ref{table:result} lists the \emph{AFT} of completing a DAG with different \emph{node\_number}. When the number of nodes is not too much, e.g., \emph{node\_number} \textless 20, the optimal solution can be obtained by exhaustive search, otherwise, it is not available due to its exponential time complexity.
The \emph{Local} here is used to indicate how the \emph{AFT} changes with executable nodes increase as a baseline.
The \emph{Remote} is better than the \emph{Local} thanks to the greater computing capability of ES. However, it is still worse than others since blind offloading leads to transmission channel and ES's processor congestion.
The \emph{RR} leverages the resources of both UE and ES, which resulting in a lower \emph{AFT} compared to the \emph{Local} and the \emph{Remote}.
The \emph{Random} may indeed find a good solution in some cases, but it is volatile and sometimes even worse than the \emph{Local} due to the inappropriate offloading, so the averaged result is. 
The \emph{HEFT} takes into account the scheduling order of nodes and further decreases \emph{AFT}. But it is still limited by the knowledge of ranking nodes.
The \emph{DTODRL} outperforms all above methods but the \emph{Optimal} and approximates the \emph{Optimal} when \emph{node\_number} \textless 20.

  

\subsection{Evaluation under Multi-user Multi-edge Setting}
When it comes to multi-user multi-edge scenario, we
look into performances of the \emph{Local}, the \emph{RR}, the \emph{Random} and the \emph{DTODRL}.
In this case, we fix $f^{ES}$ of ES with 2 processors to 10Ghz, $f^{UE}$ to 1Ghz, transmission rate between UE and ES $tr^l$ to 2 Mbps, and transmission rate between ES and ES $tr^s$ to 20 Mbps. 

We firstly discuss the results when the number of UEs changes.
Fig. \ref{subfig:3-1} $\sim$ Fig. \ref{subfig:3-4} shows that our proposed scheme surpasses others when the number of nodes of each DAG increases from 10 to 25 and the number of UE increases from 2 to 5. Generally, with the number of UEs increases, the mean \emph{AFT} of the \emph{Local} floats within a small range, and others have a raise in the mean \emph{AFT}, and the \emph{DTODRL} keeps the minimal in all cases, which illustrates that our proposed scheme is capable of multi-user multi-edge scenario.
Specially, in Fig. \ref{subfig:3-1}, the mean \emph{AFT} of the \emph{DTODRL} almost remain unchanged with the number of UEs increase. It is because that there are 3 ESs with two parallel processors, so the computation resources are saturated enough to cover all nodes that available to be offloaded in the same time. And also for this reason, in this case, the \emph{Random} has a fluctuation with the number of UEs increase compared with other settings, e.g., 20 nodes for each DAG, because it is more likely to get a better solution, avoiding the result deterioration due to unacceptable scheduling for a certain node.

Then, we fix the number of UEs to 5, and changes the number of ESs from 1 to 5.
Fig. \ref{subfig:2-1} $\sim$ Fig. \ref{subfig:2-4} shows the \emph{DTODRL} exceeds others when the number of ESs increases. The increase in the number of ESs means that more computing resources are provided to UEs, which decreases mean \emph{AFT} of UEs. In Fig. \ref{subfig:2-1}, when the number of nodes for each DAG is not very large, the reduction of the \emph{DTODRL} is relatively low, since the computation resource isn't the most crucial limitation for a lower mean \emph{AFT} in this case. In Fig. \ref{subfig:2-2} and Fig. \ref{subfig:2-3}, the reductions of the \emph{DTODRL} from 4 ESs to 5 ESs is more flat than that from 1 ESs to 4 ESs, because the computation resource almost meet the requirements for the max number of parallel executing nodes. In Fig. \ref{subfig:2-4}, the reductions of the \emph{DTODRL} maintains the same amplitude. In this case, all processors are occupied with parallel nodes and there are still more nodes waiting for the execution, so increasing ES can still have a significant effect on reducing mean \emph{AFT}.
Above results identify that our proposed scheme can be applied to complex scenarios. 



\section{Conclusion}
In this paper, we study dependent task offloading problem named DTO problem in MEC. As far as we know,
we are the first to propose a scheme to solve the general DTO problem in both single-user single-edge and multi-user multi-edge MEC scenario. We use DAG to model dependent task where nodes are subtasks and edges are their dependencies. In order to take full advantage of DAG modeling, we use GAT to capture graph structure information then concatenate it with available offloading location embedding vectors as the final state representation. To utilize GAT efficiently, we pretrain it as an encoder in unsupervised learning style in cloud layer that has massive computation resources and collected data from edges.
Inspired by some works in other graph scheduling problems \cite{HEFT,OSDI2016,sigcom}, and DRL-based task offloading \cite{dagTC2021,dagIOT2022} solution, we design a multi discrete action space DRL to adapt DTO problem scheduling process in complex offloading environment. 
The experiment results show that our proposed scheme is able to adapt simple and complex scenarios and outperforms other algorithms.

Our future work will intend to investigate more powerful 
graph-based encoders that can be embedded to MEC framework and solve the DTO problem with other optimization goals (e.g., energy consumption) under stricter resource constraints (e.g., memory, storage, and service caching).

\ifCLASSOPTIONcompsoc
  
  \section*{Acknowledgments}
\else
  \section*{Acknowledgment}
\fi

This work was supported by the National Natural Science Foundation of China Project (62172441, 62172449, 61772553), the local science and technology developing foundation guided by central government (Free exploration project 2021Szvup166), the Opening Project of State Key Laboratory of Nickel and Cobalt Resources Comprehensive Utilization(GZSYS-KY-2020-033), and the Fundamental Research Funds for the Central Universities of Central South University (2021zzts0201).

\ifCLASSOPTIONcaptionsoff
  \newpage
\fi

\bibliography{my} 
\bibliographystyle{ieeetr} 

\end{document}